\documentclass[conference]{IEEEtran}
\usepackage[T1]{fontenc}
\makeatletter\input{t1pcr.fd}\makeatother
\DeclareFontShape{T1}{pcr}{m}{sc}{<->ssub * pcr/m/n}{}
\DeclareFontShape{T1}{pcr}{b}{sc}{<->ssub * pcr/b/n}{}
\DeclareFontShape{T1}{pcr}{bx}{sc}{<->ssub * pcr/b/n}{}
\usepackage{amsmath,amssymb,mathtools}
\usepackage{booktabs}
\usepackage{array}
\usepackage{graphicx}
\usepackage{xcolor}
\usepackage{cite}
\usepackage{url}

\newcommand{\defV}[2]{\expandafter\def\csname v@#1\endcsname{#2}}
\newcommand{\V}[1]{\ifcsname v@#1\endcsname\csname v@#1\endcsname\else\textbf{\color{red}??#1??}\fi}
\defV{a100.d3x.bestm16}{0.88}
\defV{a100.d3x.eq2cost}{1.05}
\defV{a100.d3x.eq2m16}{1.26}
\defV{a100.d3x.eq2potri}{0.69}
\defV{a100.en2048.cert.n}{27}
\defV{a100.en4096.cert.n}{26}
\defV{a100.en8192.cert.n}{26}
\defV{a100.fx.eqf55.g64}{\ensuremath{2.7\times10^{-16}}}
\defV{a100.fx.eqf79.g64}{\ensuremath{2.3\times10^{-16}}}
\defV{a100.fx.eqm14.g64}{\ensuremath{2.0\times10^{-15}}}
\defV{a100.fx.eqm16.g64}{\ensuremath{2.2\times10^{-16}}}
\defV{a100.fx.f111.g64}{\ensuremath{3.3\times10^{-14}}}
\defV{a100.fx.f55.g64}{\ensuremath{6.8\times10^{-2}}}
\defV{a100.fx.f79.g64}{\ensuremath{4.1\times10^{-6}}}
\defV{a100.fx.m14.g64}{\ensuremath{7.1\times10^{-2}}}
\defV{a100.fx.m16.g64}{\ensuremath{6.7\times10^{-2}}}
\defV{a100.fx.m20.g64}{\ensuremath{4.7\times10^{-5}}}
\defV{a100.fx.native.g64}{\ensuremath{2.9\times10^{-15}}}
\defV{a100.grid.worst.ratio}{1.03}
\defV{a100.real.ssbcsstk23.blkeqgemmul8m14.sc}{\ensuremath{2.6\times10^{-13}}}
\defV{a100.realsum.air.eqm14.max}{\ensuremath{3.0\times10^{-15}}}
\defV{a100.realsum.census.eqm14.max}{\ensuremath{8.3\times10^{-15}}}
\defV{a100.realsum.eqm14.max}{138}
\defV{a100.realsum.eqm14.second}{9}
\defV{a100.realsum.ss.eqm14.max}{\ensuremath{2.6\times10^{-13}}}
\defV{a100.vs.guard.max}{1.00}
\defV{ac.n}{2048}
\defV{ac.nat.cases}{10}
\defV{ac.nat.inv}{10}
\defV{ac.s8.cases}{5}
\defV{ac.s8.inv}{5}
\defV{ac.s8.lap.g0}{\ensuremath{6.0\times10^{-14}}}
\defV{ac.s8.lap.g64}{\ensuremath{8.9\times10^{-2}}}
\defV{ac.s8.prod.g64}{\ensuremath{8.9\times10^{-2}}}
\defV{ac.s8.solve.hi}{\ensuremath{6.0\times10^{-14}}}
\defV{ac.s8.solve.lo}{\ensuremath{4.2\times10^{-14}}}
\defV{ac.up.cases}{5}
\defV{ac.up.errmax}{\ensuremath{6.0\times10^{-14}}}
\defV{ac.up.inv}{5}
\defV{adv.inj}{444}
\defV{adv.stages}{150}
\defV{ap.h100.done}{1}
\defV{ap.h100.fail.err}{\ensuremath{6.8\times10^{-2}}}
\defV{ap.h100.fail.t.ra}{\ensuremath{3.6\times10^{13}}}
\defV{ap.h100.rep.err}{\ensuremath{5.7\times10^{-16}}}
\defV{ap.l.ra.max}{520}
\defV{ap.r4090.done}{1}
\defV{ap.r4090.fail.err}{\ensuremath{7.1\times10^{-2}}}
\defV{ap.r4090.fail.t.ra}{\ensuremath{3.6\times10^{13}}}
\defV{ap.r4090.rep.err}{\ensuremath{2.1\times10^{-15}}}
\defV{apilog.default.l3}{108}
\defV{apilog.eager.l3}{108}
\defV{apilog.native.l3}{108}
\defV{as.h100.blkgemmul8m14.errmax}{\ensuremath{7.1\times10^{-2}}}
\defV{as.h100.blkscublasemufixed55.errmax}{\ensuremath{9.8\times10^{-16}}}
\defV{as.h100.blkscublasemufixed55.inv}{8}
\defV{as.h100.blkscublasemufixed55.invn}{8}
\defV{as.h100.blksgemmul8m14.errmax}{\ensuremath{5.5\times10^{-15}}}
\defV{as.h100.blksgemmul8m14.inv}{8}
\defV{as.h100.blksgemmul8m14.invn}{8}
\defV{as.r4090.blkscublasemufixed55.inv}{8}
\defV{as.r4090.blkscublasemufixed55.invn}{8}
\defV{as.r4090.blksgemmul8m14.inv}{8}
\defV{as.r4090.um.blksgemmul8m14}{\ensuremath{2.7\times10^{-15}}}
\defV{au.ra64.dev}{\ensuremath{3.5\times10^{13}}}
\defV{au.ra64.host}{\ensuremath{3.5\times10^{13}}}
\defV{au.rb.dev}{0.11}
\defV{au.rb.host}{0.12}
\defV{au.routes}{11}
\defV{aud.drawn}{6}
\defV{aud.ident}{6}
\defV{bd.all.units}{0.55}
\defV{both.fx.slope.maxdiff3}{\ensuremath{9.4\times10^{-5}}}
\defV{br30.h100.calls}{3740}
\defV{br30.h100.done}{1}
\defV{br30.h100.match}{3740}
\defV{br30.r4090.calls}{5610}
\defV{br30.r4090.done}{1}
\defV{br30.r4090.match}{5610}
\defV{cb.m.dom}{30}
\defV{cb.m.over.hi}{9,880}
\defV{cb.m.over.lo}{23}
\defV{cb.m.pairs}{30}
\defV{cb.r.blkgemmul8m14.innerall}{\ensuremath{1.2\times10^{-13}}}
\defV{cb.r.blkgemmul8m14.none}{\ensuremath{8.0\times10^{1}}}
\defV{cb.r.blkgemmul8m14.outerall}{\ensuremath{8.0\times10^{1}}}
\defV{cb.r.n}{38}
\defV{cb.r.nat.max}{\ensuremath{1.2\times10^{-13}}}
\defV{cb.r.oth.min}{\ensuremath{8.7\times10^{0}}}
\defV{ce.cw}{4}
\defV{ce.delta}{0.07}
\defV{ce.err}{0.035}
\defV{ce.full64}{2,048}
\defV{ce.kappa}{16.39}
\defV{ckh.amp.max}{33,000}
\defV{ckh.amp.min}{120}
\defV{ckh.dups}{148}
\defV{ckh.eager.worst}{\ensuremath{1.9\times10^{-7}}}
\defV{ckh.fixedeq.worst}{\ensuremath{9.0\times10^{-8}}}
\defV{ckh.lm.l3.t1.4608.blk.cublasemufixed55.dref}{\ensuremath{2.7\times10^{-10}}}
\defV{ckh.lm.l3.t1.4608.blk.gemmul8m14.dref}{\ensuremath{1.2\times10^{-9}}}
\defV{ckh.lm.l3.t1.4608.blkeq.cublasemufixed55.dref}{\ensuremath{1.3\times10^{-13}}}
\defV{ckh.lm.l3.t1.4608.blkeq.gemmul8m14.dref}{\ensuremath{4.7\times10^{-13}}}
\defV{ckh.lm.l3.t1.4608.cusolver.potri.dref}{\ensuremath{1.0\times10^{-13}}}
\defV{ckh.lm.l3.t1.4608.cusolveremueager.potri.dref}{\ensuremath{1.7\times10^{-13}}}
\defV{ckh.lm.l3.t1.4608.escale}{127}
\defV{ckh.lm.l3.t1.4608.rmse}{85,729}
\defV{ckh.lm.l3.t3.4608.blk.cublasemufixed55.dref}{\ensuremath{3.3\times10^{-8}}}
\defV{ckh.lm.l3.t3.4608.blk.gemmul8m14.dref}{\ensuremath{4.2\times10^{-8}}}
\defV{ckh.lm.l3.t3.4608.blkeq.cublasemufixed55.dref}{\ensuremath{2.3\times10^{-11}}}
\defV{ckh.lm.l3.t3.4608.blkeq.gemmul8m14.dref}{\ensuremath{2.2\times10^{-11}}}
\defV{ckh.lm.l3.t3.4608.cusolver.potri.dref}{\ensuremath{2.4\times10^{-11}}}
\defV{ckh.lm.l3.t3.4608.cusolveremueager.potri.dref}{\ensuremath{2.6\times10^{-11}}}
\defV{ckh.lm.l3.t3.4608.rmse}{101,684}
\defV{ckh.lm.l3.t6.4608.blk.cublasemufixed55.dref}{\ensuremath{1.0\times10^{-5}}}
\defV{ckh.lm.l3.t6.4608.blk.gemmul8m14.dref}{\ensuremath{1.9\times10^{-5}}}
\defV{ckh.lm.l3.t6.4608.blkeq.cublasemufixed55.dref}{\ensuremath{1.9\times10^{-8}}}
\defV{ckh.lm.l3.t6.4608.blkeq.gemmul8m14.dref}{\ensuremath{3.5\times10^{-8}}}
\defV{ckh.lm.l3.t6.4608.cusolver.potri.dref}{\ensuremath{6.5\times10^{-9}}}
\defV{ckh.lm.l3.t6.4608.cusolveremueager.potri.dref}{\ensuremath{6.4\times10^{-9}}}
\defV{ckh.lm.l3.t6.4608.rmse}{525,730}
\defV{ckh.native.worst}{\ensuremath{7.2\times10^{-8}}}
\defV{ckh.t1.cond.max}{\ensuremath{1.7\times10^{13}}}
\defV{ckh.t1.cond.min}{\ensuremath{2.2\times10^{12}}}
\defV{ckh.t1.condeq.max}{\ensuremath{1.4\times10^{5}}}
\defV{ckh.t1.fixed.worst}{\ensuremath{1.2\times10^{-8}}}
\defV{ckh.t1.gain.max}{5.1}
\defV{ckh.t1.gain.min}{-2.3}
\defV{ckh.t1.sec.max}{33,421}
\defV{ckh.t1.sec.min}{14,629}
\defV{ckh.t3.fixed.worst}{\ensuremath{6.0\times10^{-7}}}
\defV{ckh.t6.count}{8}
\defV{ckh.t6.fixed.worst}{\ensuremath{7.4\times10^{-5}}}
\defV{ckh.t6.indefinite}{5}
\defV{cv.cases}{24}
\defV{cv.t1.max}{\ensuremath{3.2\times10^{-7}}}
\defV{d3.h100.g64.blk2cublas.err}{\ensuremath{8.9\times10^{-16}}}
\defV{d3.h100.g64.blk2cublas.iqr}{0.14}
\defV{d3.h100.g64.blk2cublas.ms}{24.3}
\defV{d3.h100.g64.blk2cublas.ratio}{0.42}
\defV{d3.h100.g64.blk2eqcublasemufixed55.err}{\ensuremath{8.5\times10^{-16}}}
\defV{d3.h100.g64.blk2eqcublasemufixed55.iqr}{0.23}
\defV{d3.h100.g64.blk2eqcublasemufixed55.ms}{29.2}
\defV{d3.h100.g64.blk2eqcublasemufixed55.ratio}{0.50}
\defV{d3.h100.g64.blk2eqgemmul8m16.err}{\ensuremath{8.5\times10^{-16}}}
\defV{d3.h100.g64.blk2eqgemmul8m16.iqr}{0.11}
\defV{d3.h100.g64.blk2eqgemmul8m16.ms}{37.5}
\defV{d3.h100.g64.blk2eqgemmul8m16.ratio}{0.64}
\defV{d3.h100.g64.blkcublas.mem}{2}
\defV{d3.h100.g64.blkcublasemufixed55.err}{\ensuremath{6.8\times10^{-2}}}
\defV{d3.h100.g64.blkcublasemufixed55.iqr}{0.18}
\defV{d3.h100.g64.blkcublasemufixed55.mem}{5,474}
\defV{d3.h100.g64.blkcublasemufixed55.ms}{31.9}
\defV{d3.h100.g64.blkcublasemufixed55.ratio}{0.55}
\defV{d3.h100.g64.blkcublasemufixed55Tn.err}{\ensuremath{5.7\times10^{-16}}}
\defV{d3.h100.g64.blkcublasemufixed55Tn.iqr}{0.22}
\defV{d3.h100.g64.blkcublasemufixed55Tn.ms}{30.7}
\defV{d3.h100.g64.blkcublasemufixed55Tn.ratio}{0.53}
\defV{d3.h100.g64.blkeqgemmul8m14.err}{\ensuremath{2.1\times10^{-15}}}
\defV{d3.h100.g64.blkeqgemmul8m14.iqr}{0.099}
\defV{d3.h100.g64.blkeqgemmul8m14.ms}{38.9}
\defV{d3.h100.g64.blkeqgemmul8m14.ratio}{0.67}
\defV{d3.h100.g64.blkeqgemmul8m16.err}{\ensuremath{5.5\times10^{-16}}}
\defV{d3.h100.g64.blkeqgemmul8m16.iqr}{0.099}
\defV{d3.h100.g64.blkeqgemmul8m16.ms}{41.5}
\defV{d3.h100.g64.blkeqgemmul8m16.ratio}{0.71}
\defV{d3.h100.g64.blkgemmul8m14.err}{\ensuremath{7.1\times10^{-2}}}
\defV{d3.h100.g64.blkgemmul8m14.iqr}{0.075}
\defV{d3.h100.g64.blkgemmul8m14.mem}{3,652}
\defV{d3.h100.g64.blkgemmul8m14.ms}{37.5}
\defV{d3.h100.g64.blkgemmul8m14.ratio}{0.64}
\defV{d3.h100.g64.blkgemmul8m14Tn.err}{\ensuremath{2.1\times10^{-15}}}
\defV{d3.h100.g64.blkgemmul8m14Tn.iqr}{0.14}
\defV{d3.h100.g64.blkgemmul8m14Tn.ms}{34.9}
\defV{d3.h100.g64.blkgemmul8m14Tn.ratio}{0.60}
\defV{d3.h100.g64.blkscublasemufixed55.err}{\ensuremath{5.7\times10^{-16}}}
\defV{d3.h100.g64.blkscublasemufixed55.iqr}{0.32}
\defV{d3.h100.g64.blkscublasemufixed55.ms}{39}
\defV{d3.h100.g64.blkscublasemufixed55.ratio}{0.67}
\defV{d3.h100.g64.blksgemmul8m14.err}{\ensuremath{2.1\times10^{-15}}}
\defV{d3.h100.g64.blksgemmul8m14.iqr}{0.17}
\defV{d3.h100.g64.blksgemmul8m14.ms}{46.6}
\defV{d3.h100.g64.blksgemmul8m14.ratio}{0.80}
\defV{d3.h100.g64.cusolver.err}{\ensuremath{8.7\times10^{-16}}}
\defV{d3.h100.g64.cusolver.iqr}{0.27}
\defV{d3.h100.g64.cusolver.ms}{58.4}
\defV{d3.h100.g64.cusolver.ratio}{1.00}
\defV{d3.h100.g64.cusolveremu.err}{\ensuremath{8.7\times10^{-16}}}
\defV{d3.h100.g64.cusolveremu.iqr}{0.34}
\defV{d3.h100.g64.cusolveremu.ms}{58.4}
\defV{d3.h100.g64.cusolveremu.ratio}{1.00}
\defV{d3.h100.g64.cusolveremueager.err}{\ensuremath{8.7\times10^{-16}}}
\defV{d3.h100.g64.cusolveremueager.iqr}{0.7}
\defV{d3.h100.g64.cusolveremueager.ms}{186}
\defV{d3.h100.g64.cusolveremueager.ratio}{3.18}
\defV{d3.h100.g64.res.acc.max}{\ensuremath{5.9\times10^{-17}}}
\defV{d3.h100.g64.res.bad.min}{\ensuremath{1.9\times10^{-3}}}
\defV{d3.h100.sess.cusolver.xms.hi}{273}
\defV{d3.h100.sess.cusolver.xms.lo}{236}
\defV{d3.h100.sessions}{3}
\defV{d3.r4090.g64.blk2cublas.err}{\ensuremath{8.8\times10^{-16}}}
\defV{d3.r4090.g64.blk2cublas.iqr}{0.2}
\defV{d3.r4090.g64.blk2cublas.ms}{775}
\defV{d3.r4090.g64.blk2cublas.ratio}{1.30}
\defV{d3.r4090.g64.blk2eqcublasemufixed55.err}{\ensuremath{8.4\times10^{-16}}}
\defV{d3.r4090.g64.blk2eqcublasemufixed55.iqr}{0.15}
\defV{d3.r4090.g64.blk2eqcublasemufixed55.ms}{261}
\defV{d3.r4090.g64.blk2eqcublasemufixed55.ratio}{0.44}
\defV{d3.r4090.g64.blk2eqgemmul8m16.err}{\ensuremath{8.4\times10^{-16}}}
\defV{d3.r4090.g64.blk2eqgemmul8m16.iqr}{0.11}
\defV{d3.r4090.g64.blk2eqgemmul8m16.ms}{237}
\defV{d3.r4090.g64.blk2eqgemmul8m16.ratio}{0.40}
\defV{d3.r4090.g64.blk2eqgemmul8m16.xratio}{0.52}
\defV{d3.r4090.g64.blkcublasemufixed55.err}{\ensuremath{6.8\times10^{-2}}}
\defV{d3.r4090.g64.blkcublasemufixed55.iqr}{1.2}
\defV{d3.r4090.g64.blkcublasemufixed55.ms}{360}
\defV{d3.r4090.g64.blkcublasemufixed55.ratio}{0.61}
\defV{d3.r4090.g64.blkcublasemufixed55Tn.err}{\ensuremath{5.6\times10^{-16}}}
\defV{d3.r4090.g64.blkcublasemufixed55Tn.iqr}{0.62}
\defV{d3.r4090.g64.blkcublasemufixed55Tn.ms}{497}
\defV{d3.r4090.g64.blkcublasemufixed55Tn.ratio}{0.84}
\defV{d3.r4090.g64.blkeqgemmul8m14.err}{\ensuremath{2.1\times10^{-15}}}
\defV{d3.r4090.g64.blkeqgemmul8m14.iqr}{0.38}
\defV{d3.r4090.g64.blkeqgemmul8m14.ms}{245}
\defV{d3.r4090.g64.blkeqgemmul8m14.ratio}{0.41}
\defV{d3.r4090.g64.blkeqgemmul8m16.err}{\ensuremath{5.4\times10^{-16}}}
\defV{d3.r4090.g64.blkeqgemmul8m16.iqr}{0.43}
\defV{d3.r4090.g64.blkeqgemmul8m16.ms}{252}
\defV{d3.r4090.g64.blkeqgemmul8m16.ratio}{0.42}
\defV{d3.r4090.g64.blkgemmul8m14.err}{\ensuremath{7.1\times10^{-2}}}
\defV{d3.r4090.g64.blkgemmul8m14.iqr}{0.37}
\defV{d3.r4090.g64.blkgemmul8m14.ms}{241}
\defV{d3.r4090.g64.blkgemmul8m14.ratio}{0.40}
\defV{d3.r4090.g64.blkgemmul8m14Tn.err}{\ensuremath{2.1\times10^{-15}}}
\defV{d3.r4090.g64.blkgemmul8m14Tn.iqr}{0.51}
\defV{d3.r4090.g64.blkgemmul8m14Tn.ms}{377}
\defV{d3.r4090.g64.blkgemmul8m14Tn.ratio}{0.63}
\defV{d3.r4090.g64.blkscublasemufixed55.err}{\ensuremath{5.6\times10^{-16}}}
\defV{d3.r4090.g64.blkscublasemufixed55.iqr}{1.1}
\defV{d3.r4090.g64.blkscublasemufixed55.ms}{381}
\defV{d3.r4090.g64.blkscublasemufixed55.ratio}{0.64}
\defV{d3.r4090.g64.blksgemmul8m14.err}{\ensuremath{2.1\times10^{-15}}}
\defV{d3.r4090.g64.blksgemmul8m14.iqr}{0.17}
\defV{d3.r4090.g64.blksgemmul8m14.ms}{263}
\defV{d3.r4090.g64.blksgemmul8m14.ratio}{0.44}
\defV{d3.r4090.g64.cusolver.err}{\ensuremath{8.6\times10^{-16}}}
\defV{d3.r4090.g64.cusolver.iqr}{0.74}
\defV{d3.r4090.g64.cusolver.ms}{595}
\defV{d3.r4090.g64.cusolver.ratio}{1.00}
\defV{d3.r4090.g64.cusolveremu.err}{\ensuremath{8.7\times10^{-16}}}
\defV{d3.r4090.g64.cusolveremu.iqr}{0.8}
\defV{d3.r4090.g64.cusolveremu.ms}{548}
\defV{d3.r4090.g64.cusolveremu.ratio}{0.92}
\defV{d3.r4090.g64.cusolveremueager.err}{\ensuremath{8.7\times10^{-16}}}
\defV{d3.r4090.g64.cusolveremueager.iqr}{1.1}
\defV{d3.r4090.g64.cusolveremueager.ms}{912}
\defV{d3.r4090.g64.cusolveremueager.ratio}{1.53}
\defV{d3.r4090.g64.res.acc.max}{\ensuremath{5.9\times10^{-17}}}
\defV{d3.r4090.g64.res.bad.min}{\ensuremath{1.9\times10^{-3}}}
\defV{d3.r4090.g64.trials}{10}
\defV{d3.r4090.pending}{0}
\defV{d3.r4090.sess.blk2eqgemmul8m16.ms.hi}{244}
\defV{d3.r4090.sess.blk2eqgemmul8m16.ms.lo}{237}
\defV{d3.r4090.sess.cusolver.ms.hi}{615}
\defV{d3.r4090.sess.cusolver.ms.lo}{595}
\defV{d3.r4090.sessions}{3}
\defV{dc.g0.e99}{8}
\defV{dc.g64.e99}{67}
\defV{dc4090.above}{59}
\defV{dc4090.below}{421}
\defV{dc4090.hi}{110}
\defV{dc4090.lo}{103}
\defV{dc4090.mid}{20}
\defV{dc4090.written}{500}
\defV{dt4090.calls}{63}
\defV{dt4090.time}{63}
\defV{e2.both}{1}
\defV{e2.h100.n}{42}
\defV{e2.h100.ok}{42}
\defV{e2.r4090.n}{42}
\defV{e2.r4090.ok}{42}
\defV{ed.ratio}{620}
\defV{em.h100.n.A4}{0}
\defV{em.h100.n.captured}{3}
\defV{em.h100.n.harness}{7}
\defV{em.h100.n.timing}{9}
\defV{em.r4090.n.A4}{5}
\defV{em.r4090.n.captured}{4}
\defV{em.r4090.n.harness}{4}
\defV{em.r4090.n.timing}{7}
\defV{en.acc.n}{21}
\defV{en.asym}{0}
\defV{en.bad.cert.n}{5}
\defV{en.bfail}{0}
\defV{en.blkcublas.g0.rel}{\ensuremath{6.5\times10^{-13}}}
\defV{en.blkcublas.g32.rel}{\ensuremath{1.5\times10^{-13}}}
\defV{en.blkcublas.g64.rel}{\ensuremath{1.1\times10^{-13}}}
\defV{en.blkcublasemu.g0.rel}{\ensuremath{6.5\times10^{-13}}}
\defV{en.blkcublasemu.g32.rel}{\ensuremath{1.5\times10^{-13}}}
\defV{en.blkcublasemu.g64.rel}{\ensuremath{1.1\times10^{-13}}}
\defV{en.blkcublasemueager.g0.rel}{\ensuremath{2.3\times10^{-13}}}
\defV{en.blkcublasemueager.g32.rel}{\ensuremath{6.0\times10^{-14}}}
\defV{en.blkcublasemueager.g64.rel}{\ensuremath{6.6\times10^{-14}}}
\defV{en.blkcublasemufixed55.g0.rel}{\ensuremath{5.0\times10^{-13}}}
\defV{en.blkcublasemufixed55.g32.rel}{\ensuremath{8.5\times10^{-6}}}
\defV{en.blkcublasemufixed55.g64.rel}{n.r.b.}
\defV{en.blkcublasemufixed55Tn.g64.err}{\ensuremath{2.0\times10^{-16}}}
\defV{en.blkcublasemufixed55Tn.g64.rel}{\ensuremath{7.5\times10^{-14}}}
\defV{en.blkcublasemufixed55Xn.g64.err}{\ensuremath{6.9\times10^{-2}}}
\defV{en.blkcublasemufixed55Xn.g64.rel}{n.r.b.}
\defV{en.blkcublasemufixed79.g0.rel}{\ensuremath{2.2\times10^{-13}}}
\defV{en.blkcublasemufixed79.g32.rel}{\ensuremath{1.6\times10^{-11}}}
\defV{en.blkcublasemufixed79.g64.rel}{\ensuremath{2.6\times10^{-2}}}
\defV{en.blkeqcublasemufixed55.g0.rel}{\ensuremath{5.0\times10^{-13}}}
\defV{en.blkeqcublasemufixed55.g32.rel}{\ensuremath{1.1\times10^{-13}}}
\defV{en.blkeqcublasemufixed55.g64.rel}{\ensuremath{8.6\times10^{-14}}}
\defV{en.blkeqgemmul8m14.g0.rel}{\ensuremath{2.9\times10^{-12}}}
\defV{en.blkeqgemmul8m14.g32.rel}{\ensuremath{6.5\times10^{-13}}}
\defV{en.blkeqgemmul8m14.g64.rel}{\ensuremath{5.0\times10^{-13}}}
\defV{en.blkgemmul8m14.g0.rel}{\ensuremath{2.9\times10^{-12}}}
\defV{en.blkgemmul8m14.g32.rel}{\ensuremath{3.1\times10^{-5}}}
\defV{en.blkgemmul8m14.g64.rel}{n.r.b.}
\defV{en.blkgemmul8m14Tn.g64.err}{\ensuremath{9.9\times10^{-16}}}
\defV{en.blkgemmul8m14Tn.g64.rel}{\ensuremath{1.5\times10^{-13}}}
\defV{en.blkgemmul8m14Xn.g64.err}{\ensuremath{7.1\times10^{-2}}}
\defV{en.blkgemmul8m14Xn.g64.rel}{n.r.b.}
\defV{en.blkgemmul8m16.g0.rel}{\ensuremath{3.0\times10^{-13}}}
\defV{en.blkgemmul8m16.g32.rel}{\ensuremath{2.0\times10^{-7}}}
\defV{en.blkgemmul8m16.g64.rel}{n.r.b.}
\defV{en.blkgemmul8m20.g0.rel}{\ensuremath{1.5\times10^{-13}}}
\defV{en.blkgemmul8m20.g32.rel}{\ensuremath{4.8\times10^{-11}}}
\defV{en.blkgemmul8m20.g64.rel}{n.r.b.}
\defV{en.cert.n}{26}
\defV{en.emu.cases}{7}
\defV{en.emu.over.hi}{402}
\defV{en.emu.over.lo}{40}
\defV{en.frame.native.g0}{\ensuremath{6.5\times10^{-13}}}
\defV{en.frame.native.g32}{\ensuremath{6.5\times10^{-13}}}
\defV{en.frame.native.g64}{\ensuremath{6.5\times10^{-13}}}
\defV{en.h.tn.max}{\ensuremath{1.5\times10^{-13}}}
\defV{en.hyb.n}{4}
\defV{en.leaf.max}{\ensuremath{2.7\times10^{-15}}}
\defV{en.leaf.nu}{\ensuremath{4.6\times10^{-16}}}
\defV{en.n}{4096}
\defV{en.pairs}{30}
\defV{en.rem.big.max}{272}
\defV{en.rem.max}{1.8}
\defV{en.rem.thr}{\ensuremath{1.0\times10^{-6}}}
\defV{en.sec.maxdiff}{\ensuremath{3.9\times10^{-9}}}
\defV{en.sec.n}{34}
\defV{en.sec.ok}{34}
\defV{en.self.cases}{66}
\defV{en.self.magmax}{0.2}
\defV{en.self.ok}{66}
\defV{en.up.all}{0.12}
\defV{en.up.lauum}{0.033}
\defV{en.up.t}{0.053}
\defV{en.up.x}{0.12}
\defV{en.vac.n}{4}
\defV{en.vac.over.hi}{\ensuremath{3.6\times10^{17}}}
\defV{en.vac.over.lo}{1.3}
\defV{en4.bfail}{0}
\defV{en4.cert.n}{26}
\defV{en4.sec.n}{34}
\defV{en4.sec.ok}{34}
\defV{en4.up.lauum}{0.032}
\defV{en4.up.t}{0.05}
\defV{en4.up.x}{0.11}
\defV{en4.v2}{1}
\defV{eqv.inner}{0}
\defV{eqv.innern}{160}
\defV{eqv.outer}{224}
\defV{eqv.outern}{224}
\defV{eqv.pipe.prod}{0}
\defV{eqv.pipe.solve}{5}
\defV{eqv.pipe.solven}{5}
\defV{fair.h100.emutwo}{1.20}
\defV{fair.r4090.solve}{1.07}
\defV{fair.r4090.tn}{1.54}
\defV{gr.aflip}{0}
\defV{gr.aflip2}{0}
\defV{gr.allow.a}{\ensuremath{4.9\times10^{-4}}}
\defV{gr.bflip}{0}
\defV{gr.bflip2}{0}
\defV{gr.blkcublas.bg}{\ensuremath{1.1\times10^{-1}}}
\defV{gr.blkcublas.g0.at}{\ensuremath{1.5\times10^{-2}}}
\defV{gr.blkcublas.g0.err}{\ensuremath{2.3\times10^{-15}}}
\defV{gr.blkcublas.g32.at}{\ensuremath{1.5\times10^{-2}}}
\defV{gr.blkcublas.g32.err}{\ensuremath{2.2\times10^{-15}}}
\defV{gr.blkcublas.g64.at}{\ensuremath{1.5\times10^{-2}}}
\defV{gr.blkcublas.g64.err}{\ensuremath{2.0\times10^{-15}}}
\defV{gr.blkcublasemu.bg}{\ensuremath{1.1\times10^{-1}}}
\defV{gr.blkcublasemu.g0.at}{\ensuremath{1.5\times10^{-2}}}
\defV{gr.blkcublasemu.g0.err}{\ensuremath{2.3\times10^{-15}}}
\defV{gr.blkcublasemu.g32.at}{\ensuremath{1.5\times10^{-2}}}
\defV{gr.blkcublasemu.g32.err}{\ensuremath{2.2\times10^{-15}}}
\defV{gr.blkcublasemu.g64.at}{\ensuremath{1.5\times10^{-2}}}
\defV{gr.blkcublasemu.g64.err}{\ensuremath{2.0\times10^{-15}}}
\defV{gr.blkcublasemueager.bg}{\ensuremath{7.1\times10^{-3}}}
\defV{gr.blkcublasemueager.g0.at}{\ensuremath{7.2\times10^{-3}}}
\defV{gr.blkcublasemueager.g0.err}{\ensuremath{2.1\times10^{-15}}}
\defV{gr.blkcublasemueager.g32.at}{\ensuremath{9.1\times10^{-3}}}
\defV{gr.blkcublasemueager.g32.err}{\ensuremath{2.0\times10^{-15}}}
\defV{gr.blkcublasemueager.g64.at}{\ensuremath{1.3\times10^{-2}}}
\defV{gr.blkcublasemueager.g64.err}{\ensuremath{2.0\times10^{-15}}}
\defV{gr.blkcublasemufixed55.bg}{\ensuremath{1.1\times10^{-2}}}
\defV{gr.blkcublasemufixed55.g0.at}{\ensuremath{4.3\times10^{0}}}
\defV{gr.blkcublasemufixed55.g0.err}{\ensuremath{7.3\times10^{-16}}}
\defV{gr.blkcublasemufixed55.g32.at}{\ensuremath{5.6\times10^{8}}}
\defV{gr.blkcublasemufixed55.g32.err}{\ensuremath{2.3\times10^{-8}}}
\defV{gr.blkcublasemufixed55.g64.at}{\ensuremath{3.5\times10^{13}}}
\defV{gr.blkcublasemufixed55.g64.err}{\ensuremath{6.9\times10^{-2}}}
\defV{gr.blkcublasemufixed79.bg}{\ensuremath{9.1\times10^{-3}}}
\defV{gr.blkcublasemufixed79.g0.at}{\ensuremath{8.7\times10^{-3}}}
\defV{gr.blkcublasemufixed79.g0.err}{\ensuremath{3.3\times10^{-16}}}
\defV{gr.blkcublasemufixed79.g32.at}{\ensuremath{3.0\times10^{1}}}
\defV{gr.blkcublasemufixed79.g32.err}{\ensuremath{1.4\times10^{-15}}}
\defV{gr.blkcublasemufixed79.g64.at}{\ensuremath{2.5\times10^{11}}}
\defV{gr.blkcublasemufixed79.g64.err}{\ensuremath{3.1\times10^{-6}}}
\defV{gr.blkeqcublasemufixed55.bg}{\ensuremath{1.0\times10^{-2}}}
\defV{gr.blkeqcublasemufixed55.g0.at}{\ensuremath{4.3\times10^{0}}}
\defV{gr.blkeqcublasemufixed55.g0.err}{\ensuremath{7.3\times10^{-16}}}
\defV{gr.blkeqcublasemufixed55.g32.at}{\ensuremath{4.3\times10^{0}}}
\defV{gr.blkeqcublasemufixed55.g32.err}{\ensuremath{2.5\times10^{-16}}}
\defV{gr.blkeqcublasemufixed55.g64.at}{\ensuremath{4.3\times10^{0}}}
\defV{gr.blkeqcublasemufixed55.g64.err}{\ensuremath{2.0\times10^{-16}}}
\defV{gr.blkeqgemmul8m14.bg}{\ensuremath{5.1\times10^{-2}}}
\defV{gr.blkeqgemmul8m14.g0.at}{\ensuremath{2.8\times10^{1}}}
\defV{gr.blkeqgemmul8m14.g0.err}{\ensuremath{3.0\times10^{-15}}}
\defV{gr.blkeqgemmul8m14.g32.at}{\ensuremath{2.8\times10^{1}}}
\defV{gr.blkeqgemmul8m14.g32.err}{\ensuremath{1.1\times10^{-15}}}
\defV{gr.blkeqgemmul8m14.g64.at}{\ensuremath{2.8\times10^{1}}}
\defV{gr.blkeqgemmul8m14.g64.err}{\ensuremath{1.0\times10^{-15}}}
\defV{gr.blkgemmul8m14.bg}{\ensuremath{5.3\times10^{-2}}}
\defV{gr.blkgemmul8m14.g0.at}{\ensuremath{2.8\times10^{1}}}
\defV{gr.blkgemmul8m14.g0.err}{\ensuremath{3.0\times10^{-15}}}
\defV{gr.blkgemmul8m14.g32.at}{\ensuremath{1.1\times10^{9}}}
\defV{gr.blkgemmul8m14.g32.err}{\ensuremath{8.1\times10^{-8}}}
\defV{gr.blkgemmul8m14.g64.at}{\ensuremath{3.5\times10^{13}}}
\defV{gr.blkgemmul8m14.g64.err}{\ensuremath{7.1\times10^{-2}}}
\defV{gr.blkgemmul8m16.bg}{\ensuremath{3.9\times10^{-3}}}
\defV{gr.blkgemmul8m16.g0.at}{\ensuremath{1.0\times10^{-1}}}
\defV{gr.blkgemmul8m16.g0.err}{\ensuremath{3.1\times10^{-16}}}
\defV{gr.blkgemmul8m16.g32.at}{\ensuremath{3.0\times10^{6}}}
\defV{gr.blkgemmul8m16.g32.err}{\ensuremath{3.8\times10^{-10}}}
\defV{gr.blkgemmul8m16.g64.at}{\ensuremath{3.5\times10^{13}}}
\defV{gr.blkgemmul8m16.g64.err}{\ensuremath{6.7\times10^{-2}}}
\defV{gr.blkgemmul8m20.bg}{\ensuremath{3.9\times10^{-3}}}
\defV{gr.blkgemmul8m20.g0.at}{\ensuremath{3.9\times10^{-3}}}
\defV{gr.blkgemmul8m20.g0.err}{\ensuremath{3.1\times10^{-16}}}
\defV{gr.blkgemmul8m20.g32.at}{\ensuremath{1.5\times10^{2}}}
\defV{gr.blkgemmul8m20.g32.err}{\ensuremath{1.2\times10^{-14}}}
\defV{gr.blkgemmul8m20.g64.at}{\ensuremath{1.7\times10^{12}}}
\defV{gr.blkgemmul8m20.g64.err}{\ensuremath{2.3\times10^{-5}}}
\defV{gr.bmax}{0.12}
\defV{gr.fixed.xmax}{\ensuremath{1.2\times10^{1}}}
\defV{gr.lauum.products}{30}
\defV{gr.n}{4096}
\defV{gr.native.amax}{\ensuremath{8.7\times10^{-2}}}
\defV{gr.pred.count}{9}
\defV{gr.pred.hi}{18}
\defV{gr.pred.lo}{0.11}
\defV{gr.trtri.products}{900}
\defV{gr.up.lauum}{0.019}
\defV{gr.up.lauum.n}{30}
\defV{gr.up.t}{0.052}
\defV{gr.up.t.n}{450}
\defV{gr.up.x}{0.12}
\defV{gr.up.x.n}{450}
\defV{gv.reldiff}{\ensuremath{8.7\times10^{-16}}}
\defV{h100.ctl.cond_1e8.chol}{\ensuremath{5.4\times10^{-9}}}
\defV{h100.ctl.cond_1e8.chol.eager}{\ensuremath{5.4\times10^{-9}}}
\defV{h100.cw.worst.ratio}{1.56}
\defV{h100.default.calls.all}{1090}
\defV{h100.default.emulated.all}{0}
\defV{h100.err.worst.ratio}{1.05}
\defV{h100.fx.eqf55.g64}{\ensuremath{2.9\times10^{-16}}}
\defV{h100.fx.eqf55.slope}{0.00}
\defV{h100.fx.eqf79.g64}{\ensuremath{2.4\times10^{-16}}}
\defV{h100.fx.eqf79.slope}{0.00}
\defV{h100.fx.eqm14.g64}{\ensuremath{2.0\times10^{-15}}}
\defV{h100.fx.eqm14.slope}{0.00}
\defV{h100.fx.eqm16.g64}{\ensuremath{2.3\times10^{-16}}}
\defV{h100.fx.eqm16.slope}{0.00}
\defV{h100.fx.f111.g64}{\ensuremath{3.3\times10^{-14}}}
\defV{h100.fx.f111.slope}{0.00}
\defV{h100.fx.f55.g64}{\ensuremath{6.8\times10^{-2}}}
\defV{h100.fx.f55.slope}{0.81}
\defV{h100.fx.f79.g64}{\ensuremath{4.1\times10^{-6}}}
\defV{h100.fx.f79.slope}{0.96}
\defV{h100.fx.m14.g64}{\ensuremath{7.1\times10^{-2}}}
\defV{h100.fx.m14.slope}{0.76}
\defV{h100.fx.m16.g64}{\ensuremath{6.7\times10^{-2}}}
\defV{h100.fx.m16.slope}{0.87}
\defV{h100.fx.m20.g64}{\ensuremath{4.7\times10^{-5}}}
\defV{h100.fx.m20.slope}{0.96}
\defV{h100.fx.native.g64}{\ensuremath{2.9\times10^{-15}}}
\defV{h100.fx.native.slope}{0.00}
\defV{h100.fx.raw.invariant}{1}
\defV{h100.spy.calls}{220}
\defV{h100.v132.cublas}{13.3.0}
\defV{h100.v132.cusolver}{12.1.0}
\defV{h100.v132.err.reldiff.pct}{1}
\defV{h100k.csfx.b55.potri.g64}{\ensuremath{1.2\times10^{-2}}}
\defV{h100k.d3x.bestm16}{0.64}
\defV{h100k.d3x.eq2cost}{1.06}
\defV{h100k.d3x.eq2f55}{1.13}
\defV{h100k.d3x.eq2m16}{1.46}
\defV{h100k.d3x.eq2potri}{0.44}
\defV{h100k.d3x.eqf55}{0.92}
\defV{h100k.d3x.lauum}{0.70}
\defV{h100k.en2048.cert.n}{27}
\defV{h100k.loo.blkcublasemufixed55.n4096.lo}{\ensuremath{3.7\times10^{-3}}}
\defV{h100k.loo.blkcublasemufixed55.n8192.lo}{\ensuremath{1.9\times10^{-3}}}
\defV{h100k.loo.blkgemmul8m14.n4096.lo}{\ensuremath{3.7\times10^{-3}}}
\defV{h100k.loo.blkgemmul8m14.n8192.lo}{\ensuremath{1.9\times10^{-3}}}
\defV{h100k.n16k.blkcublas.g64}{\ensuremath{4.8\times10^{-15}}}
\defV{h100k.n16k.blkcublasemufixed55.g64}{\ensuremath{7.1\times10^{-2}}}
\defV{h100k.n16k.blkcublasemufixed79.g64}{\ensuremath{5.9\times10^{-6}}}
\defV{h100k.n16k.blkeqcublasemufixed55.g64}{\ensuremath{3.5\times10^{-16}}}
\defV{h100k.n16k.blkeqgemmul8m14.g64}{\ensuremath{3.9\times10^{-15}}}
\defV{h100k.n16k.blkgemmul8m14.g64}{\ensuremath{7.4\times10^{-2}}}
\defV{h100k.n16k.blkgemmul8m16.g64}{\ensuremath{7.0\times10^{-2}}}
\defV{h100k.n16k.blkgemmul8m20.g64}{\ensuremath{8.3\times10^{-5}}}
\defV{h100k.s8.blkeqcublasemufixed55.g64.hi}{\ensuremath{2.9\times10^{-16}}}
\defV{h100k.s8.blkeqcublasemufixed55.g64.lo}{\ensuremath{2.7\times10^{-16}}}
\defV{h100k.s8.blkgemmul8m14.g64.hi}{\ensuremath{7.6\times10^{-2}}}
\defV{h100k.s8.blkgemmul8m14.g64.lo}{\ensuremath{7.1\times10^{-2}}}
\defV{h100k.vs.guard.max}{1.00}
\defV{h100k.vs.nver}{9}
\defV{h100k.vs.v137027.blkeqcublasemufixed55.g64}{\ensuremath{1.3\times10^{-14}}}
\defV{h100k.vs.v13804.blkeqcublasemufixed55.g64}{\ensuremath{2.9\times10^{-16}}}
\defV{kern.4096.s44.bits}{103}
\defV{kern.4096.s44.ratio}{1.69}
\defV{kern.4096.s48.bits}{107}
\defV{kern.4096.s48.ratio}{2.17}
\defV{ko.4096.s44.int8us}{2,350}
\defV{ko.4096.s48.int8us}{47.2}
\defV{l40s.csfx.b111.potri.g64}{\ensuremath{9.8\times10^{-16}}}
\defV{l40s.csfx.b55.potri.g64}{\ensuremath{1.2\times10^{-2}}}
\defV{l40s.csfx.b55.potrieq.g64}{\ensuremath{6.7\times10^{-16}}}
\defV{l40s.csfx.b79.potri.g64}{\ensuremath{1.7\times10^{-7}}}
\defV{l40s.d3x.bestm16}{0.39}
\defV{l40s.d3x.eq2cost}{1.01}
\defV{l40s.d3x.eq2m16}{0.30}
\defV{l40s.d3x.eq2potri}{1.29}
\defV{l40s.en2048.cert.n}{27}
\defV{l40s.en4096.cert.n}{26}
\defV{l40s.en8192.cert.n}{26}
\defV{l40s.fx.eqf55.g64}{\ensuremath{2.9\times10^{-16}}}
\defV{l40s.fx.eqf79.g64}{\ensuremath{2.4\times10^{-16}}}
\defV{l40s.fx.eqm14.g64}{\ensuremath{2.0\times10^{-15}}}
\defV{l40s.fx.eqm16.g64}{\ensuremath{2.3\times10^{-16}}}
\defV{l40s.fx.f111.g64}{\ensuremath{3.3\times10^{-14}}}
\defV{l40s.fx.f55.g64}{\ensuremath{6.8\times10^{-2}}}
\defV{l40s.fx.f79.g64}{\ensuremath{4.1\times10^{-6}}}
\defV{l40s.fx.m14.g64}{\ensuremath{7.1\times10^{-2}}}
\defV{l40s.fx.m16.g64}{\ensuremath{6.7\times10^{-2}}}
\defV{l40s.fx.m20.g64}{\ensuremath{4.7\times10^{-5}}}
\defV{l40s.fx.native.g64}{\ensuremath{2.9\times10^{-15}}}
\defV{l40s.grid.worst.ratio}{1.02}
\defV{l40s.real.ssbcsstk23.blkcublas.sc}{\ensuremath{2.5\times10^{-15}}}
\defV{l40s.real.ssbcsstk23.blkcublasemufixed55.sc}{\ensuremath{5.1\times10^{-14}}}
\defV{l40s.real.ssbcsstk23.blkeqcublasemufixed55.sc}{\ensuremath{5.4\times10^{-15}}}
\defV{l40s.real.ssbcsstk23.blkeqgemmul8m14.sc}{\ensuremath{2.7\times10^{-13}}}
\defV{l40s.real.ssbcsstk23.blkeqgemmul8m16.sc}{\ensuremath{2.5\times10^{-15}}}
\defV{l40s.real.ssbcsstk23.blkgemmul8m14.sc}{\ensuremath{1.7\times10^{-12}}}
\defV{l40s.real.ssbcsstk23.cusolveremueagerpotri.sc}{\ensuremath{3.1\times10^{-15}}}
\defV{l40s.realsum.air.eager.max}{\ensuremath{9.9\times10^{-16}}}
\defV{l40s.realsum.air.eqf55.max}{\ensuremath{7.9\times10^{-16}}}
\defV{l40s.realsum.air.eqm14.max}{\ensuremath{3.0\times10^{-15}}}
\defV{l40s.realsum.air.eqm16.max}{\ensuremath{5.9\times10^{-16}}}
\defV{l40s.realsum.air.f55.max}{\ensuremath{6.3\times10^{-14}}}
\defV{l40s.realsum.air.m14.max}{\ensuremath{1.6\times10^{-13}}}
\defV{l40s.realsum.air.native.max}{\ensuremath{1.4\times10^{-15}}}
\defV{l40s.realsum.census.eager.max}{\ensuremath{1.7\times10^{-15}}}
\defV{l40s.realsum.census.eqf55.max}{\ensuremath{2.3\times10^{-15}}}
\defV{l40s.realsum.census.eqm14.max}{\ensuremath{8.4\times10^{-15}}}
\defV{l40s.realsum.census.eqm16.max}{\ensuremath{1.2\times10^{-15}}}
\defV{l40s.realsum.census.f55.max}{\ensuremath{3.7\times10^{-12}}}
\defV{l40s.realsum.census.m14.max}{\ensuremath{1.5\times10^{-11}}}
\defV{l40s.realsum.census.native.max}{\ensuremath{2.9\times10^{-15}}}
\defV{l40s.realsum.eqf55.max}{2.2}
\defV{l40s.realsum.eqm14.arg}{bcsstk23}
\defV{l40s.realsum.eqm14.max}{108}
\defV{l40s.realsum.eqm14.second}{10}
\defV{l40s.realsum.eqm16.max}{1.0}
\defV{l40s.realsum.guard.max}{1.8}
\defV{l40s.realsum.n}{24}
\defV{l40s.realsum.ss.eager.max}{\ensuremath{1.2\times10^{-14}}}
\defV{l40s.realsum.ss.eqf55.max}{\ensuremath{3.4\times10^{-14}}}
\defV{l40s.realsum.ss.eqm14.max}{\ensuremath{2.7\times10^{-13}}}
\defV{l40s.realsum.ss.eqm16.max}{\ensuremath{2.9\times10^{-14}}}
\defV{l40s.realsum.ss.f55.max}{\ensuremath{3.2\times10^{-13}}}
\defV{l40s.realsum.ss.m14.max}{\ensuremath{1.7\times10^{-12}}}
\defV{l40s.realsum.ss.native.max}{\ensuremath{3.6\times10^{-14}}}
\defV{l40s.spy.default.written}{53}
\defV{l40s.spy.eager.written}{155}
\defV{l40s.vs.guard.max}{1.01}
\defV{l40s.vs.nver}{9}
\defV{l40s.vs.v13602.blkeqcublasemufixed55.g64}{\ensuremath{1.3\times10^{-14}}}
\defV{l40s.vs.v137027.blkeqcublasemufixed55.g64}{\ensuremath{2.9\times10^{-16}}}
\defV{ld.differ}{219}
\defV{ld.entries}{540}
\defV{ld.hosts}{8}
\defV{ld.jobs}{18}
\defV{ld.jobs.entries}{9,720}
\defV{ld.jobs.seq}{9,720}
\defV{ld.seq}{540}
\defV{ld.worst}{0.37}
\defV{lk.capped.rel}{\ensuremath{1.4\times10^{-13}}}
\defV{lk.fresh.rel}{\ensuremath{6.0\times10^{-14}}}
\defV{ob.cases}{8}
\defV{ob.t1.fixed}{\ensuremath{1.3\times10^{-9}}}
\defV{ob.t1.fixed.base}{\ensuremath{1.5\times10^{-9}}}
\defV{ob.t1.rmse}{85,725}
\defV{ob.t1.rmse.base}{85,729}
\defV{ob.t3.fixedeq}{\ensuremath{3.0\times10^{-11}}}
\defV{ord.g64.blocks.8192.cublasemueager}{\ensuremath{3.5\times10^{-15}}}
\defV{ord.g64.blocks.8192.gemmul8m16}{\ensuremath{9.9\times10^{-4}}}
\defV{ord.g64.sorted.8192.cublasemueager}{\ensuremath{4.9\times10^{-15}}}
\defV{ord.g64.sorted.8192.gemmul8m16}{\ensuremath{5.5\times10^{-11}}}
\defV{pf.checks}{6}
\defV{pj.ckhlml3t1.4608.blkcublas.epsx}{\ensuremath{4.0\times10^{-15}}}
\defV{pj.ckhlml3t1.4608.blkcublas.varreal}{\ensuremath{1.3\times10^{-11}}}
\defV{pj.ckhlml3t1.4608.blkcublas.varrel}{\ensuremath{4.1\times10^{-12}}}
\defV{pj.ckhlml3t1.4608.blkgemmul8m14.epsx}{\ensuremath{5.7\times10^{-11}}}
\defV{pj.ckhlml3t1.4608.blkgemmul8m14.varreal}{\ensuremath{4.0\times10^{-14}}}
\defV{pj.ckhlml3t1.4608.blkgemmul8m14.varrel}{\ensuremath{2.1\times10^{-10}}}
\defV{pj.ckhlml3t1.4608.kappav.max}{9.8}
\defV{pj.ckhlml3t3.4608.blkcublas.varrel}{\ensuremath{8.6\times10^{-8}}}
\defV{pj.ckhlml3t3.4608.blkgemmul8m14.varrel}{\ensuremath{1.6\times10^{-8}}}
\defV{pj.ckhlml3t3.4608.kappav.max}{\ensuremath{9.7\times10^{2}}}
\defV{pj.ckhlml3t6.4608.cusolverpotri.varrel}{\ensuremath{6.0\times10^{-2}}}
\defV{pj.ckhlml3t6.4608.kappav.max}{\ensuremath{9.5\times10^{5}}}
\defV{pk.default.all}{830}
\defV{pk.default.emu}{0}
\defV{pr.h100.eqcost.mhi}{1.043}
\defV{pr.h100.eqcost.mlo}{1.036}
\defV{pr.h100.nodes}{2}
\defV{pr.h100.nresolved}{7}
\defV{pr.h100.sessions}{3}
\defV{pr.r4090.nresolved}{7}
\defV{r4090.adp.bits.eq.hi}{75}
\defV{r4090.adp.bits.g0.hi}{75}
\defV{r4090.adp.bits.g0.lo}{68}
\defV{r4090.adp.calls}{636}
\defV{r4090.ctl.cond_1e8.chol}{\ensuremath{5.3\times10^{-9}}}
\defV{r4090.ctl.cond_1e8.chol.eager}{\ensuremath{5.3\times10^{-9}}}
\defV{r4090.ctl.gp_opt.chol}{\ensuremath{3.1\times10^{-7}}}
\defV{r4090.cublas}{13.8.1}
\defV{r4090.cw.worst.ratio}{1.95}
\defV{r4090.default.calls.all}{1090}
\defV{r4090.err.worst.ratio}{1.02}
\defV{r4090.fit.slope}{1.00}
\defV{r4090.fx.eqf55.g64}{\ensuremath{2.7\times10^{-16}}}
\defV{r4090.fx.eqf55.um}{\ensuremath{4.5\times10^{-16}}}
\defV{r4090.fx.eqf79.g64}{\ensuremath{2.3\times10^{-16}}}
\defV{r4090.fx.eqf79.um}{\ensuremath{2.7\times10^{-16}}}
\defV{r4090.fx.eqm14.g64}{\ensuremath{2.0\times10^{-15}}}
\defV{r4090.fx.eqm14.um}{\ensuremath{2.8\times10^{-15}}}
\defV{r4090.fx.eqm16.g64}{\ensuremath{2.1\times10^{-16}}}
\defV{r4090.fx.eqm16.um}{\ensuremath{2.5\times10^{-16}}}
\defV{r4090.fx.f111.g64}{\ensuremath{3.3\times10^{-14}}}
\defV{r4090.fx.f111.um}{\ensuremath{2.6\times10^{-16}}}
\defV{r4090.fx.f55.g64}{\ensuremath{6.8\times10^{-2}}}
\defV{r4090.fx.f55.um}{\ensuremath{2.3\times10^{-5}}}
\defV{r4090.fx.f79.g64}{\ensuremath{4.1\times10^{-6}}}
\defV{r4090.fx.f79.um}{\ensuremath{1.4\times10^{-12}}}
\defV{r4090.fx.m14.g64}{\ensuremath{7.1\times10^{-2}}}
\defV{r4090.fx.m14.um}{\ensuremath{1.1\times10^{-4}}}
\defV{r4090.fx.m16.g64}{\ensuremath{6.7\times10^{-2}}}
\defV{r4090.fx.m16.um}{\ensuremath{6.1\times10^{-7}}}
\defV{r4090.fx.m20.g64}{\ensuremath{4.7\times10^{-5}}}
\defV{r4090.fx.m20.um}{\ensuremath{1.9\times10^{-11}}}
\defV{r4090.fx.native.g64}{\ensuremath{2.8\times10^{-15}}}
\defV{r4090.fx.native.um}{\ensuremath{3.4\times10^{-15}}}
\defV{r4090.inv.eager.count}{15}
\defV{r4090.inv.native.count}{30}
\defV{r4090.probe.s0.bits}{62}
\defV{r4090.probe.s200.bits}{257}
\defV{r4090.probe.tri_8192.time}{1.16}
\defV{r4090.spy.eager.syrk.calls}{63}
\defV{r4090.spy.eager.trmm.hi}{125}
\defV{r4090.spy.eager.trsm.hi}{129}
\defV{r4090.spy.eager.trsm.lo}{66}
\defV{r4090.step.small.above}{109}
\defV{r4090.step.small.below}{107}
\defV{r4090.step.small.t48}{5.59}
\defV{r4090.step.small.t52}{6.51}
\defV{r4090.thr.2048.cap}{110}
\defV{r4090.thr.2048.captime.lo}{1.60}
\defV{r4090.thr.2048.emu}{107}
\defV{r4090.thr.2048.emutime.hi}{0.98}
\defV{r4090.thr.4096.cap}{107}
\defV{r4090.thr.4096.emu}{103}
\defV{rb.cases}{24}
\defV{rb.delta}{\ensuremath{2.7\times10^{-13}}}
\defV{rb.lratio}{0.142}
\defV{rb.mu}{0.142}
\defV{rb.shacases}{16}
\defV{rb.shamatch}{16}
\defV{rb.t1.fixedcert}{\ensuremath{7.9\times10^{-7}}}
\defV{rb.t1.max}{\ensuremath{7.8\times10^{-7}}}
\defV{rb.t3.fixedcert}{\ensuremath{7.4\times10^{-3}}}
\defV{rb.t3.max}{\ensuremath{7.4\times10^{-3}}}
\defV{rb.t6.over}{8}
\defV{rc.cases}{12}
\defV{rc.eta.max}{\ensuremath{4.8\times10^{-19}}}
\defV{rc.steps}{4}
\defV{ref.dps}{60}
\defV{ref.nmax}{512}
\defV{ref.ref.cw.max}{\ensuremath{1.9\times10^{-11}}}
\defV{ref.ref.norm.max}{\ensuremath{1.2\times10^{-15}}}
\defV{rm.air12vmm341.span}{20}
\defV{rm.bcsstk23.kappa}{\ensuremath{3.9\times10^{7}}}
\defV{rm.bcsstk23.span}{35}
\defV{rm.cal7vmm585.span}{32}
\defV{rm.n.hi}{8190}
\defV{rm.n.lo}{1074}
\defV{rm.span.hi}{44}
\defV{rm.span.lo}{20}
\defV{rm.ss.count}{16}
\defV{rm.ss.span.hi}{44}
\defV{rm.ss.span.lo}{23}
\defV{rp.blkcublasemufixed55.4096.bottom16}{\ensuremath{6.8\times10^{-2}}}
\defV{rp.blkcublasemufixed55.4096.lauum_only}{\ensuremath{6.9\times10^{-2}}}
\defV{rp.blkcublasemufixed55.4096.none}{\ensuremath{6.9\times10^{-2}}}
\defV{rp.blkcublasemufixed55.4096.top16}{\ensuremath{2.0\times10^{-16}}}
\defV{rp.blkcublasemufixed55.8192.bottom16}{\ensuremath{6.8\times10^{-2}}}
\defV{rp.blkcublasemufixed55.8192.lauum_only}{\ensuremath{6.8\times10^{-2}}}
\defV{rp.blkcublasemufixed55.8192.none}{\ensuremath{6.8\times10^{-2}}}
\defV{rp.blkcublasemufixed55.8192.top16}{\ensuremath{1.6\times10^{-2}}}
\defV{rp.blkgemmul8m14.4096.bottom16}{\ensuremath{7.0\times10^{-2}}}
\defV{rp.blkgemmul8m14.4096.exin.hi}{71.5}
\defV{rp.blkgemmul8m14.4096.exin.lo}{69.5}
\defV{rp.blkgemmul8m14.4096.exout.hi}{8.7}
\defV{rp.blkgemmul8m14.4096.lauum_only}{\ensuremath{7.1\times10^{-2}}}
\defV{rp.blkgemmul8m14.4096.none}{\ensuremath{7.1\times10^{-2}}}
\defV{rp.blkgemmul8m14.4096.top16}{\ensuremath{9.9\times10^{-16}}}
\defV{rp.blkgemmul8m14.8192.bottom16}{\ensuremath{7.1\times10^{-2}}}
\defV{rp.blkgemmul8m14.8192.lauum_only}{\ensuremath{7.1\times10^{-2}}}
\defV{rp.blkgemmul8m14.8192.none}{\ensuremath{7.1\times10^{-2}}}
\defV{rp.blkgemmul8m14.8192.top16}{\ensuremath{1.6\times10^{-2}}}
\defV{rp.full.count}{10}
\defV{rp.full.max}{\ensuremath{2.0\times10^{-15}}}
\defV{rp.partial.count}{64}
\defV{rp.partial.min}{\ensuremath{1.6\times10^{-2}}}
\defV{rp2.blkcublasemufixed55.4096.inner_all}{\ensuremath{2.0\times10^{-16}}}
\defV{rp2.blkcublasemufixed55.4096.outer_all}{\ensuremath{6.9\times10^{-2}}}
\defV{rp2.blkcublasemufixed55.8192.inner_all}{\ensuremath{2.9\times10^{-16}}}
\defV{rp2.blkcublasemufixed55.8192.outer_all}{\ensuremath{6.8\times10^{-2}}}
\defV{rp2.blkgemmul8m14.4096.inner_all}{\ensuremath{9.9\times10^{-16}}}
\defV{rp2.blkgemmul8m14.4096.outer_all}{\ensuremath{7.1\times10^{-2}}}
\defV{rp2.blkgemmul8m14.8192.inner_all}{\ensuremath{2.0\times10^{-15}}}
\defV{rp2.blkgemmul8m14.8192.outer_all}{\ensuremath{7.1\times10^{-2}}}
\defV{seed.slope.maxspread}{0.01}
\defV{seed4090.slope.maxspread}{0.01}
\defV{spyc.eager.calls}{108}
\defV{tm4090.8192.scaleg0.emu}{0.737}
\defV{tm4090.8192.scaleg64.emu}{0.918}
\defV{tm4090.8192.scaleg64.emueagereq}{1.050}
\defV{tr.kern.n}{3}
\defV{tr.kern.same}{3}
\defV{tr.trsm.n}{4}
\defV{tr.trsm.native}{4}
\defV{um.diag.log2}{80}
\defV{um.g}{40}
\defV{v132.thr.time.hi}{148}
\defV{v132.thr.time.lo}{16}
\defV{vec.n}{94}
\defV{vec.problems}{0}
\defV{wind.done}{1}
\defV{wind.n}{5}
\defV{wind.status}{5}
\defV{x4090.csfx.b55.potri.g64}{\ensuremath{1.2\times10^{-2}}}
\defV{x4090.d3x.bestm16}{0.40}
\defV{x4090.d3x.eq2cost}{1.01}
\defV{x4090.d3x.eq2f55}{0.33}
\defV{x4090.d3x.eq2m16}{0.30}
\defV{x4090.d3x.eq2potri}{1.31}
\defV{x4090.d3x.eqf55}{0.27}
\defV{x4090.d3x.lauum}{0.57}
\defV{x4090.en2048.cert.n}{27}
\defV{x4090.en8192.cert.n}{26}

\urldef{\kFPtc}\path{cublasLt_imma_emu::cutlass_80_tensorop_d884gemm_128x128_16x3_nn_align1}
\urldef{\kNvjet}\path{nvjet_sm90_biu_256x128_128x4_2x1_v_bz_coopA_smod_TNN}
\urldef{\kProlog}\path{prolog_mod_dfma<oz2_int8_dgemm,...>}
\urldef{\kCsv}\path{kernels_h100_*/k_*.csv}
\urldef{\kNative}\path{sm90_xmma_gemm_f64f64_f64f64_f64_nn_n_tilesize128x64x16_...}

\newtheorem{proposition}{Proposition}
\newtheorem{theorem}{Theorem}

\begin{document}

\title{Accurate Products, Inaccurate Inverse:\\ Emulated FP64 BLAS in Blocked Triangular Inversion}
\author{\IEEEauthorblockN{Erzheng Zhang}\IEEEauthorblockA{Department of Computer Science, University of Toronto\\Toronto, Canada\\erwin@cs.toronto.edu}}
\maketitle

\begin{abstract}
FP64 emulation on integer tensor units now ships in cuBLAS and GEMMul8, and each emulated matrix product can be
checked against an accuracy grade. We ask if a routine built from such products inherits their accuracy, and show
that blocked triangular inversion, the core of LAPACK's inverse of a symmetric positive definite matrix, does not.
Under fixed-precision emulation every product passes the grade-B error bound of Demmel et al., yet the inverse errs at
\V{gr.blkgemmul8m14.g64.err}, against \V{gr.blkcublas.g64.err} natively. The matrices are well conditioned after
diagonal scaling, but the scaling spans 64 binades. The cause is one product, $T = X_{22}L_{21}$, whose summed index
carries both the diagonal scaling and its inverse, which a grade-B bound does not absorb. Computing those products
natively restores native accuracy; leaving any one of them emulated does not. The pattern holds on four GPUs, and on
real covariance matrices the fixed-precision routes lose up to about four orders of accuracy. Equilibration, a solve-based
association and cuBLAS's automatic dynamic precision avoid the loss, and an exact error recurrence audits a completed
run. On an RTX 4090 the fastest accurate emulated inversion takes \V{d3.r4090.g64.blk2eqgemmul8m16.ratio} of
cuSOLVER's time.
Where each product's scaling falls decides if per-product accuracy carries over to a composite routine, and that is
what builders of emulated libraries need to check.
\end{abstract}

\section{Introduction}
\label{sec:intro}

\begin{figure*}[t]
\centering
\includegraphics[width=\textwidth]{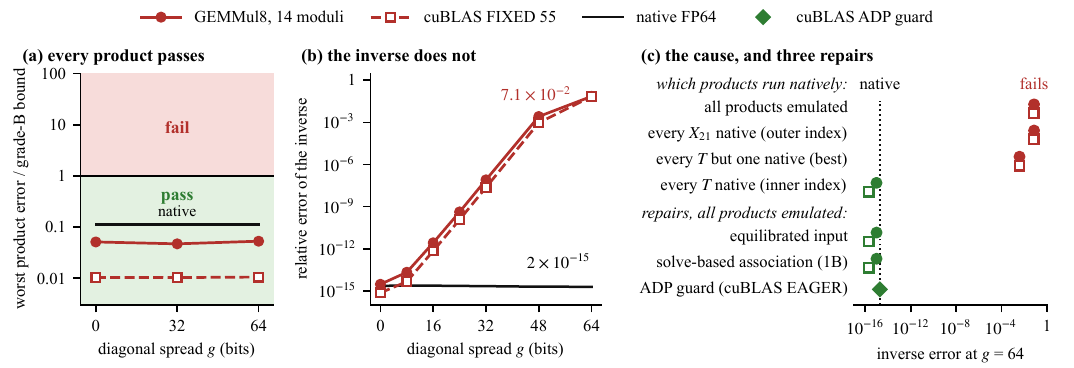}
\caption{Every product passes; the inverse does not (H100, $n = \V{gr.n}$). (a) Worst TRTRI product error over its
grade-B bound; (b) the inverse's error; (c) at $g = 64$, native $T$ restores it, native $X_{21}$ does not, and three
repairs restore it with every product emulated.}
\label{fig:main}
\end{figure*}

Since CUDA 13, cuBLAS can run DGEMM, TRMM and TRSM on integer tensor units by emulating FP64
arithmetic~\cite{schwarz2026,cublasdoc}. A guard, automatic dynamic precision (ADP), estimates per call how many bits a
product needs. GEMMul8 offers fixed-precision emulation~\cite{ozaki2026,uchino2025sc}, and emulation has been proposed
as a default algorithm for a wide range of applications~\cite{uchino2026isc}. Demmel et al.\ grade each emulated
product by the error bound it attains~\cite{demmel2026grade}. LAPACK-level libraries such as cuSOLVER, MAGMA and SLATE
will call these products, so their accuracy is the accuracy that matters.

A composite routine combines many products. Demmel et al.\ proved that grade-B products keep tiled Cholesky
componentwise accurate. LAPACK's inverse of a symmetric positive definite matrix has a second half: triangular
inversion (TRTRI), then the product of the inverse factor with its transpose (LAUUM). Whether it inherits the
products' accuracy has not been asked.

It does not. On an H100 at $n = \V{gr.n}$, every product of the blocked inversion meets the row-and-column-scaled
(B-type) error test with room to spare; the largest ratio, \V{gr.bmax} of the allowance, is a native product's. Yet
through fixed-precision
emulation the inverse is wrong in its second digit: relative error \V{gr.blkgemmul8m14.g64.err}, against
\V{gr.blkcublas.g64.err} for native FP64 on the same well-conditioned input. The cause is not a defective product but
which index of a product the diagonal scaling falls on (Fig.~\ref{fig:main}).

Users meet such matrices whenever variables carry different units, as covariances do. They use explicit inverses
where many inverse entries are needed, as in kriging variances~\cite{wackernagel2003}. A covariance with unit factors
from $10^{-6}$ to $10^{6}$ loses accuracy the same way (14 moduli, RTX 4090, $n = 8192$: \V{r4090.fx.m14.um} against
\V{r4090.fx.native.um}, Appendix~\ref{app:tables}).

The contributions are four.
\begin{itemize}
\item \emph{Mechanism} (Section~\ref{sec:mechanism}): a three-by-three example proving that grade B permits the loss,
  an exact error recurrence with an a posteriori enclosure, and a provably scale-equivariant association.
\item \emph{Diagnosis} (Sections~\ref{sec:grade}, \ref{sec:replace} and~\ref{sec:breadth}): on four GPUs and two libraries, every product
  passes while the inverse fails; a causal test needs every inner-index product.
\item \emph{Repairs and cost} (Sections~\ref{sec:repair}--\ref{sec:cost}): three repairs that keep every product
  emulated, an enclosure that certifies
  every accurate run, and complete device-resident timings.
\item \emph{Application} (Section~\ref{sec:app}): the loss reaches cokriging predictions on census data, within a
  certified bound at the model's nugget.
\end{itemize}

\section{Background and related work}
\label{sec:background}

\subsection{FP64 emulation and its precision demand}
The Ozaki scheme~\cite{ozaki2012,ootomo2024} computes an FP64 product $C = AB$ on integer units. It scales each row
of $A$ and each column of $B$ by a power of two, so that the largest entry of each has a fixed exponent. It splits the
scaled entries into INT8 slices and multiplies slice pairs exactly with integer accumulation on tensor cores. The
partial products are summed in FP64. Ozaki-II replaces the slices by residues modulo $q$ coprime moduli and
reconstructs the product by the Chinese remainder theorem~\cite{ozaki2026}. More slices or moduli buy more bits at
more cost.

In both, the bits available to $c_{ij}$ are set relative to $r_is_j$, with $r_i$ and $s_j$ the largest entries of row
$i$ of $|A|$ and column $j$ of $|B|$~\cite{ootomo2024,uchino2025,uchino2025sc}. An entry much smaller than $r_is_j$
loses about $\log_2(r_is_j/|c_{ij}|)$ bits, its \emph{scale excess}, so a fixed bit count suffices only when every
entry's excess is small. Fasi, Mary and Mikaitis bound the fixed-point width needed for a target precision and show
that ADP's bit estimate can be very pessimistic~\cite{fasi2026}. cuBLAS runs either ADP or FIXED, a set number of bits~\cite{cublasdoc}. ADP estimates the bits a call
needs, then emulates with that many bits or dispatches to native DGEMM; its strategy, DEFAULT or EAGER, sets how
readily emulation is chosen. GEMMul8 takes the number of moduli
$q$. All of this is stated for one product; this paper asks what happens when a routine combines many.

\subsection{Accuracy grades}
Demmel et al.~\cite{demmel2026grade} grade a GEMM by the bound it attains:
\begin{itemize}
\item grade A, $|\widehat C - C| \le f(k)\,u\,|A||B|$ componentwise, the bound of classical multiplication;
\item grade B, $|\widehat C - C| \le f(k)\,u\,\mathrm{rowmax}(|A|)\,\mathrm{colmax}(|B|)$, the natural model of row-
  and column-scaled emulation;
\item grade C, the normwise bound that suffices for LU and QR~\cite{demmel1992,demmel2007}.
\end{itemize}
Grade B is invariant under row scaling of $A$ and column scaling of $B$, not under scaling of the summed index.
This paper uses
one normalization, $f(k) = k$: with $r_X$ the largest ratio of a product's error to $u$ times the grade-$X$ right-hand
side, the product passes the \emph{A-type} or \emph{B-type test} when $r_X/k \le 1$. Since $(|A||B|)_{ij} \le
k\,r_is_j$, the two tests are not nested. The test $r_B/k \le 1$ is stricter than Demmel et al.'s own benchmark
threshold $f_B(k) < k^2$. A quantity is \emph{certified} when a bound with every rounding enclosed proves it.

\subsection{Blocked triangular inversion and LAUUM}
LAPACK inverts a symmetric positive definite $A$ by Cholesky (\texttt{potrf}, $A = LL^T$), triangular inversion
(\texttt{trtri}, $X = L^{-1}$) and LAUUM ($A^{-1} = X^TX$), together \texttt{potri}. Recursive blocked TRTRI splits $L
= \begin{psmallmatrix} L_{11} & \\ L_{21} & L_{22}\end{psmallmatrix}$, inverts the diagonal blocks, $X_{ii} =
L_{ii}^{-1}$, and forms
\begin{equation}
T = X_{22} L_{21}, \qquad X_{21} = -\,T X_{11}.
\label{eq:split}
\end{equation}
We call $T$ the \emph{inner-index} product and $X_{21}$ the \emph{outer-index} product. In Du Croz and Higham's
terms~\cite{ducroz1992}, \eqref{eq:split} is Method~2B. Method~2C forms $T$ and then solves with $L_{11}$ (the
reference \texttt{dtrtri}); Method~1B forms $W = L_{21}X_{11}$ and solves with $L_{22}$, $X_{21} = -L_{22}^{-1}W$. In
ordinary arithmetic 1B and 2C keep a componentwise residual bound and 2B does not.

\subsection{Related work}
Mixed-precision solvers are judged by solver-level outcomes, often with iterative refinement as the
remedy~\cite{carson2018,haidar2018,higham2022}; LU and HPL on INT8 emulation fail the residual check at too few
splits~\cite{luszczek2025}. Block algorithms on fast level-3 BLAS stay normwise backward stable but may lose
componentwise accuracy and invariance under power-of-radix scaling~\cite{demmel1992}. Carson and Ma analyse block LU
with an Ozaki trailing update and a row- and column-norm bound~\cite{carson2026}; they do not treat inversion or
diagonal scaling. Diagonal scaling of a single product's operands is studied for fast multiplication~\cite{ballard2016},
and equilibration is classical~\cite{vandersluis1969,lapack}. Known, then, are that operand scaling consumes a
fixed-precision product's bits~\cite{fasi2026,ootomo2024}, that equilibration helps, and that the association of a
block inversion matters in ordinary arithmetic~\cite{ducroz1992}. To Demmel et al., whose theorem concerns Cholesky,
this paper adds a composite in which grade-B products do not suffice. To Du Croz and Higham it adds an exposure of
$X_{22}L_{21}$ that a diagonal scaling creates and native FP64 does not see. To the single-product evaluations of
ADP~\cite{schwarz2026} and the width bounds of Fasi et al.~\cite{fasi2026} it adds the behaviour inside a complete
shipped routine and the cost of each repair.

\section{Why per-product checks do not compose}
\label{sec:mechanism}

\begin{figure}[t]
\centering
\includegraphics[width=\columnwidth]{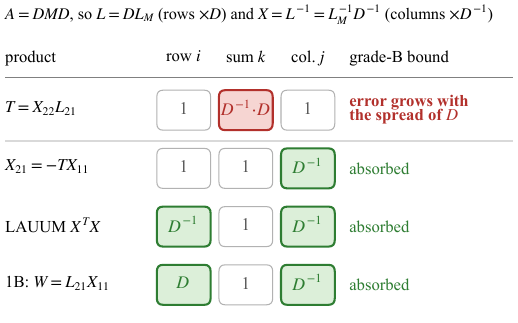}
\caption{Only $T$ sums over a scaled index (red); grade B absorbs the scaling on row $i$ and column $j$ (green).}
\label{fig:scaling}
\end{figure}

\paragraph{Where the scaling falls}
Let $A = DMD$ with $D = \mathrm{diag}(2^{k_i})$, $k_i \in \{0,\dots,g\}$, and $M = L_ML_M^T$. Then $L = DL_M$ and
$L^{-1} = L_M^{-1}D^{-1}$. Fig.~\ref{fig:scaling} shows which scaling each product's row, summed and column index
carries. In $T = X_{22}L_{21}$ the summed index $k$ carries $d_k^{-1}$ on the columns of $X_{22}$ and $d_k$ on the rows
of $L_{21}$. $T = L_{M,22}^{-1}L_{M,21}$ itself does not depend on $D$, but a row of $X_{22}$ or a column of $L_{21}$
spans as many binades as the exponents in the contracted block. The span is $g$ only if both extremes fall in one
block, so it depends on the order of the $k_i$. $X_{21}$ and LAUUM's $X^TX = D^{-1}L_M^{-T}L_M^{-1}D^{-1}$ carry $D$ on
outer indices only; Method~1B forms no product that sums over a scaled index. Native FP64 is unaffected: native
inversion by~\eqref{eq:split} is \emph{bit-for-bit scale equivariant} in every test here, that is, its computed
inverse of $DMD$ equals $D^{-1}$ times its computed inverse of $M$ times $D^{-1}$ exactly. The loss needs a product
bound that sees the inner scaling.

\paragraph{A grade-B bound does not secure the inverse}
An \emph{admissible evaluation} of the recursion returns, for each product, any value whose error meets the stated
bound, and evaluates every other operation exactly.

\begin{proposition}
\label{prop:cex}
Let $L_M$ be the $3 \times 3$ lower-triangular matrix of ones, $D = \mathrm{diag}(1, 1, 2^g)$ and $L = DL_M$, split
after the first row and column. For every $\delta$ with $|\delta| \le f_B(2)\,u\,2^g$ there is an admissible evaluation
of the blocked inverse under grade B in which $\|\widehat X - L^{-1}\|_F / \|L^{-1}\|_F = |\delta| / (4 +
4^{-g})^{1/2}$, while $\kappa_2(M) \approx \V{ce.kappa}$ for every $g$.
\end{proposition}

\begin{IEEEproof}
$X_{11} = 1$, $X_{22} = \begin{psmallmatrix} 1 & 0 \\ -1 & 2^{-g} \end{psmallmatrix}$ and $L_{21} = (1, 2^g)^T$, so
$T = (1, 0)^T$, whose second entry is the cancellation $-1 + 2^{-g}2^g$. For it the row maximum of $|X_{22}|$ is 1,
the column maximum of $|L_{21}|$ is $2^g$ and $k = 2$, so $\widehat T_2 = \delta$ meets grade B. With every other
operation exact, $\widehat X$ differs from $L^{-1}$ only in entry $(3,1)$, by $-\delta$, and $\|L^{-1}\|_F^2 = 4 +
4^{-g}$.
\end{IEEEproof}

With $f(k) = k$ the bound admits a relative error of about $2^{g-53}$, \V{ce.full64} at $g = 64$, while grade A admits %
only $O(u)$ on the same entry (Appendix~\ref{app:full}). The example applies verbatim to Method~2C, since solving
with $L_{11} = 1$ changes nothing. Tiled Cholesky escapes this because the right-hand sides of its grade-B TRSM and SYRK
are bounded through the factor's row norms, at most $(a_{ii}a_{jj})^{1/2}$, which carry the scaling on outer
indices~\cite{demmel2026grade}. The inversion product $T$ has no such protection: its grade-B allowance grows with the
spread of the contracted block.

\paragraph{How product errors reach the inverse}
$L$ is the stored factor; $X = L^{-1}$ and $Y = X^TX = (LL^T)^{-1}$ are exact, and the factorization's error is
outside the statement. At a split with column block $\rho_1$ and row block $\rho_2$, write the computed products as
$\widehat T = \widehat X_{22}L_{21} + E_s$ and $\widehat X_{21} = -(\widehat T\widehat X_{11} + F_s)$, and let $\Phi_s
= -(E_s\widehat X_{11} + F_s)$, $\Delta = \widehat X - X$ and $\Delta_\ell$ the error of leaf $\ell$'s inverse. LAUUM
returns $\widehat Y = \mathrm{fl}(\mathrm{sym}(\widehat X^T\widehat X + G))$ with output error $G$ and $\mathrm{sym}(C)
= (C + C^T)/2$, one rounded addition per entry.

\begin{theorem}
\label{thm:comp}
Let the negations and block copies be exact and every addition and halving of the symmetrization finite (round to
nearest; underflow allowed).

(a) At every split
\begin{multline}
\Delta_{21} = \Phi_s - \Delta_{22}L_{21}\widehat X_{11} - (\widehat T - E_s)\Delta_{11}\\ + \Delta_{22}L_{21}\Delta_{11},
\label{eq:rec}
\end{multline}
and $\Delta = \sum_\nu \Lambda_\nu\Psi_\nu\Gamma_\nu$ over the splits and leaves $\nu$, with $\Lambda_\nu =
X_{:,\rho_2}L_{22}$, $\Gamma_\nu = L_{11}X_{\rho_1,:}$, $\Psi_s = \Phi_s - \Delta_{22}L_{21}\Delta_{11}$ and $\Psi_\ell
= \Delta_\ell$ (a leaf's block is both $\rho_1$ and $\rho_2$).

(b) If $|E_s| \le \mathcal E_s$, $|F_s| \le \mathcal F_s$, $|\Delta_\ell| \le B_\ell$ and $|G| \le \mathcal G$, then
$|\Delta| \le B$ for $B$ assembled from the leaves upward by
\begin{multline}
B_{21} = \mathcal E_s|\widehat X_{11}| + \mathcal F_s + B_{22}|L_{21}\widehat X_{11}|\\
 + (|\widehat T| + \mathcal E_s)B_{11} + B_{22}|L_{21}|B_{11},
\label{eq:B}
\end{multline}
and $|\widehat Y - Y| \le |\widehat X|^TB + B^T|\widehat X| + B^TB + \mathrm{sym}(\mathcal G) + 2u|\widehat Y| +
2^{-1074}\mathbf 1\mathbf 1^T$. %

(c) Let $A = DMD$, $L = DL_M$, and write $B' = BD$, $\widehat X' = \widehat XD$ ($M$'s frame). If every $T$ meets grade
A with $\mathcal E_s = c\,u|\widehat X_{22}||L_{21}|$, the other products meet grade B, and $c$ and $B_\ell D_\ell$ do
not depend on $D$, then \eqref{eq:B} for $B'$ and the resulting bound on $D|\widehat Y - Y|D$ are one function of the
run's $\widehat X'$, $\widehat T$, $D\widehat YD$, $L_M$ and $c$ for every power-of-two $D$, except that the absolute
term becomes $2^{-1074}dd^T$ ($d$ the diagonal of $D$). If $T$ meets grade B only, $\mathcal E_s|\widehat X_{11}|$ %
carries at most the spread $\rho_s = \max_{k\in\rho_2}d_k / \min_{k\in\rho_2}d_k$ of $D$ over the contracted block.
\end{theorem}

No error is assumed small, and no exact or high-precision inverse enters: the enclosure is computed from $L$, the
computed $\widehat X$ and $\widehat T$, and per-product allowances. Part (c) is the mechanism in one line: grade A on
$T$ keeps the bound free of $D$, grade B lets it grow with $\rho_s$, and Proposition~\ref{prop:cex} attains that
growth. The proof is in Appendix~\ref{app:full}.

\paragraph{An association that keeps the scaling outside}
Method~1B forms $W = L_{21}X_{11} = D_2L_{M,21}L_{M,11}^{-1}D_1^{-1}$ and solves $X_{21} = -L_{22}^{-1}W$ recursively
down to native TRSM leaves, so no product contracts a scale; per split of order $2h$ it spends $3h^3$ flops against the
product form's $4h^3$. Call a GEMM \emph{outer-equivariant} if $G(D_1A, BD_2) = D_1G(A, B)D_2$ exactly for power-of-two
diagonal $D_1$, $D_2$.

\begin{proposition}
\label{prop:solve}
Under hypotheses (i) to (v) of Appendix~\ref{app:full}, the solve-based inverse of $DMD$ equals $D^{-1}$ times that of
$M$ times $D^{-1}$, bit for bit.
\end{proposition}

Equivariance transports the error with the result; it does not imply accuracy, which is measured. The product form
has no such property, and neither has LAPACK's Method~2C on a lower factor, which forms the same $X_{22}L_{21}$: on a
transparent CPU emulator (8 slices, $n = \V{ac.n}$) with only that product emulated, its error grows from
\V{ac.s8.lap.g0} at $g = 0$ to \V{ac.s8.lap.g64} at $g = 64$, as the product form's does (Appendix~\ref{app:tables}).

\section{Experimental setup}
\label{sec:setup}

Table~\ref{tab:setup} lists the design. Two implementations run the same split rule and call structure, form every
product with $\beta = 0$ and perform each update as a separate operation. The numerical harness takes LAPACK's host
factor, inverts the leaves on the host and sends every GEMM to the device; all accuracy results and grades use it.
Grades cover every output entry of every product, against a long-double reference product whose own error is bounded
entry by entry (Appendix~\ref{app:grading}). The device-resident implementation factors with cuSOLVER and solves the
leaves with \path{cublasDtrsm}; all times use it. Each timed route runs from the uploaded matrix to the inverse,
including \texttt{potrf} and any equilibration, \V{d3.r4090.g64.trials} times interleaved with the other routes in a
rotating order, in several sessions per GPU. Its products are not graded; an audit regrades the products of the timed
routes on the RTX 4090 and ties the two implementations together (Appendix~\ref{app:setup}).
\texttt{CUBLAS\_EMULATION\_STRATEGY} was unset, so DEFAULT is the library's own default.

The reference for $DMD$ is $D^{-1}X_MD^{-1}$, with $X_M$ LAPACK's inverse of $M$, checked against
high-precision inverses (Appendix~\ref{app:setup}). The error of an inverse is $\epsilon_X = \|\widehat X -
X_{\mathrm{ref}}\|_F/\|X_{\mathrm{ref}}\|_F$; for the real matrices it is measured in the equilibrated frame ($\widehat X$ scaled by
$\mathrm{diag}(A)^{1/2}$ on both sides), where the normwise error cannot hide the small-diagonal block. The \emph{slope} is the number of bits of $\epsilon_X$ lost per
unit of $g$. Routes are named by their GEMM back end (Table~\ref{tab:setup}); ``eq.''\ marks the power-of-two
equilibrated input, and unless a row says otherwise every GEMM of the blocked inversion runs on that back end. For a
group of matrices a table gives the worst $\epsilon_X$ in the group.

\begin{table}[t]
\caption{Experimental design.}
\label{tab:setup}
\centering\footnotesize
\setlength{\tabcolsep}{3pt}
\begin{tabular}{@{}>{\raggedright\arraybackslash}p{0.2\columnwidth}>{\raggedright\arraybackslash}p{0.77\columnwidth}@{}} %
\toprule
factor & values \\
\midrule
synthetic inputs & $DMD$, $M = W^TW/n + I$, $W$ standard normal ($\kappa(M) \approx 6$); $k_i$ uniform on $\{0,\dots,g\}$,
$g \in \{0, 8, 16, 24, 32, 48, 64\}$; $n \in \{2048, 4096, 8192\}$, breadth runs at 16384; seeds 0--4 \\[1pt]
controls & exponents sorted or in contiguous chunks; $\kappa = 10^8$, $10^{14}$ with uniform diagonal; Gaussian-process
kernel; unit factors $10^{-6}$ to $10^{6}$ (diagonal span $2^{\V{um.diag.log2}}$) \\[1pt]
real matrices & \V{rm.ss.count} SuiteSparse SPD matrices and 8 raw-unit covariances (census, air quality), $n$
\V{rm.n.lo} to \V{rm.n.hi}, diagonal spans $2^{\V{rm.span.lo}}$ to $2^{\V{rm.span.hi}}$ \\[1pt]
application & cokriging, California housing, $n \in \{4608, 9216\}$, nugget $\tau \in \{10^{-1}, 10^{-3},
10^{-6}\}$ \\[1pt]
GPUs and libraries & FP64-scarce: RTX 4090, L40S; FP64-rich: H100, A100. cuBLAS 13.8.1, cuSOLVER 12.3.5, GEMMul8
v3.5.4; ten releases, cuBLAS 13.0 to cuBLAS 13.10, on H100, L40S and A100 \\[1pt]
GEMM back ends & \emph{native} FP64 DGEMM; ADP with DEFAULT or EAGER; F55, F79, F111: FIXED with that many bits;
$q$ = 14, 16, 20: GEMMul8 moduli \\[1pt]
routes & cuSOLVER \texttt{potri}; blocked product form; hybrids with every $T$ or every $X_{21}$ native; solve-based
(Method~1B); each also equilibrated; LAUUM as one GEMM or blocked; leaves of width 256 \\
\bottomrule
\end{tabular}
\end{table}

\section{Results}
\label{sec:results}

\subsection{Does the inverse fail although every product passes?}
\label{sec:grade}
\label{sec:fixed}

Grading every output entry of every TRTRI product of ten routes at three spreads answers this. Every product of every
route passes the B-type test (Fig.~\ref{fig:main}(a)), yet the inverses of the fixed-precision
routes fail (Fig.~\ref{fig:main}(b)). A product-level check therefore cannot tell these routes from accurate ones. A
normwise audit and a reference-free residual agree, so the failure is not an artifact of the grading or the reference
(Appendix~\ref{app:tables}).

\subsection{Is the inner index the cause?}
\label{sec:replace}

If the mechanism is right, the inner-index products carry the loss and the outer-index products do not. Rerunning 14
moduli and FIXED 55 at $g = 64$ with chosen TRTRI products computed by native DGEMM tests this
(Fig.~\ref{fig:main}(c)). Computing every $T$ natively restores native accuracy, at a cost in speed
(Appendix~\ref{app:cost}); computing every $X_{21}$ natively changes nothing, and leaving a single $T$ emulated still
leaves at least \V{h100k.loo.blkgemmul8m14.n4096.lo}. Over every replacement plan tested (Appendix~\ref{app:tables}),
the \V{rp.partial.count} that leave seven or more $T$ emulated err at \V{rp.partial.min} or more, and the
\V{rp.full.count} that compute every $T$ natively at \V{rp.full.max} or less. Each $T$ contributes through its own
contracted block, so the repair needs every one of them.

\subsection{What repairs it?}
\label{sec:repair}
\label{sec:adp}
\label{sec:default}

Anything that moves the scaling off the summed index should repair the inverse. Power-of-two equilibration does:
inverting $SAS$, with $S$ the power-of-two diagonal that brings every diagonal entry into $[1, 4)$, and scaling the
result by $S$ again removes $D$ up to a factor of two per entry, exactly. Every equilibrated fixed route keeps its
$g = 0$ accuracy at every $g$ (Table~\ref{tab:general}). It needs only the diagonal, which every SPD inversion has, and
no change to the library.

\begin{table}[t]
\caption{$\epsilon_X$ at $g = 64$, $n = 8192$ (real: L40S, worst per group). Fixed routes fail; equilibration
repairs all but bcsstk23.}
\label{tab:general}
\centering\scriptsize
\setlength{\tabcolsep}{1pt}
\begin{tabular*}{\columnwidth}{@{\extracolsep{\fill}}lccccc@{}}
\toprule
 & & & & \multicolumn{2}{c}{equilibrated} \\
\cmidrule(lr){5-6} %
input & native & $q$=14 & F55 & $q$=14 & F55 \\
\midrule
H100 & \V{h100.fx.native.g64} & \V{h100.fx.m14.g64} & \V{h100.fx.f55.g64} & \V{h100.fx.eqm14.g64} & \V{h100.fx.eqf55.g64} \\
A100 & \V{a100.fx.native.g64} & \V{a100.fx.m14.g64} & \V{a100.fx.f55.g64} & \V{a100.fx.eqm14.g64} & \V{a100.fx.eqf55.g64} \\
RTX 4090 & \V{r4090.fx.native.g64} & \V{r4090.fx.m14.g64} & \V{r4090.fx.f55.g64} & \V{r4090.fx.eqm14.g64} & \V{r4090.fx.eqf55.g64} \\
L40S & \V{l40s.fx.native.g64} & \V{l40s.fx.m14.g64} & \V{l40s.fx.f55.g64} & \V{l40s.fx.eqm14.g64} & \V{l40s.fx.eqf55.g64} \\
\midrule
census & \V{l40s.realsum.census.native.max} & \V{l40s.realsum.census.m14.max} & \V{l40s.realsum.census.f55.max} & \V{l40s.realsum.census.eqm14.max} & \V{l40s.realsum.census.eqf55.max} \\
SuiteSparse & \V{l40s.realsum.ss.native.max} & \V{l40s.realsum.ss.m14.max} & \V{l40s.realsum.ss.f55.max} & \V{l40s.realsum.ss.eqm14.max} & \V{l40s.realsum.ss.eqf55.max} \\
\bottomrule
\end{tabular*}
\end{table}

The solve-based association repairs the inverse without touching the input. It is bit-for-bit scale equivariant in
every tested case, as Proposition~\ref{prop:solve} predicts, and its error on the raw input never exceeds
\V{as.h100.blksgemmul8m14.errmax} (Appendix~\ref{app:tables}).

ADP avoids the loss with no change to the input or the algorithm: every guarded route stays within a factor
\V{h100.err.worst.ratio} of native over every GPU's synthetic grid. Under DEFAULT the choice depends on the GPU: inside cuSOLVER's \texttt{potri} it emulates no call on the FP64-rich GPUs and some on the
FP64-scarce ones (Appendix~\ref{app:switch}).

\subsection{Why do the repairs work?}
\label{sec:why}

Theorem~\ref{thm:comp}(c) names the links: the size of $T$'s miss of grade A, the spread $\rho_s$ behind it, and the
bits $T$ is given. Four measurements test them.

\paragraph{The miss on $T$ sets the error}
At $g = 64$ every $T$ has a 99th-percentile scale excess $e_{99}$ of \V{rp.blkgemmul8m14.4096.exin.lo} to
\V{rp.blkgemmul8m14.4096.exin.hi} bits, against at most \V{rp.blkgemmul8m14.4096.exout.hi} for $X_{21}$, so a fixed
budget leaves $T$ short and nothing else. The A-type test sees this (Table~\ref{tab:miss}): only the fixed routes miss
it on $T$, by a factor that grows with $g$, and where that factor is large the inverse's error follows $u$ times it to
within about one order of magnitude (Appendix~\ref{app:tables}). The size of the miss, not the miss itself, decides: at
$g = 0$ the fixed routes miss grade A and still invert accurately.

\begin{table}[t]
\caption{Only the fixed routes miss the A-type test on $T$ by a factor that grows with $g$ (H100, $n = \V{gr.n}$).}
\label{tab:miss}
\centering\footnotesize
\setlength{\tabcolsep}{3pt}
\begin{tabular*}{\columnwidth}{@{\extracolsep{\fill}}lccc@{}}
\toprule
 & \multicolumn{3}{c}{$r_A/k$ on $T$} \\
\cmidrule(lr){2-4} %
route & $g = 0$ & $g = 32$ & $g = 64$ \\
\midrule
native & \V{gr.blkcublas.g0.at} & \V{gr.blkcublas.g32.at} & \V{gr.blkcublas.g64.at} \\
EAGER & \V{gr.blkcublasemueager.g0.at} & \V{gr.blkcublasemueager.g32.at} & \V{gr.blkcublasemueager.g64.at} \\
\midrule
$q$=14 & \V{gr.blkgemmul8m14.g0.at} & \V{gr.blkgemmul8m14.g32.at} & \V{gr.blkgemmul8m14.g64.at} \\
F55 & \V{gr.blkcublasemufixed55.g0.at} & \V{gr.blkcublasemufixed55.g32.at} & \V{gr.blkcublasemufixed55.g64.at} \\
\midrule
$q$=14, eq. & \V{gr.blkeqgemmul8m14.g0.at} & \V{gr.blkeqgemmul8m14.g32.at} & \V{gr.blkeqgemmul8m14.g64.at} \\
F55, eq. & \V{gr.blkeqcublasemufixed55.g0.at} & \V{gr.blkeqcublasemufixed55.g32.at} & \V{gr.blkeqcublasemufixed55.g64.at} \\
\bottomrule
\end{tabular*}
\end{table}

\paragraph{The spread that matters is per block}
Reordering the same exponents changes $\rho_s$ and nothing else. With the exponents sorted, so that each contracted
block spans few binades, 16 moduli give \V{ord.g64.sorted.8192.gemmul8m16} at $n = 8192$, $g = 64$, against
\V{h100.fx.m16.g64} in random order (contiguous chunks: Appendix~\ref{app:tables}).

\paragraph{Equilibration removes the spread}
After equilibration the diagonal spans less than a factor of four at every $g$, so no block carries a spread that
grows with $g$. The equilibrated routes' miss on $T$ at $g = 64$ stays at its $g = 0$ size (Table~\ref{tab:miss}),
and ADP's estimates on the equilibrated input stay at most \V{r4090.adp.bits.eq.hi} bits.

\begin{figure}[t]
\centering
\includegraphics[width=\columnwidth]{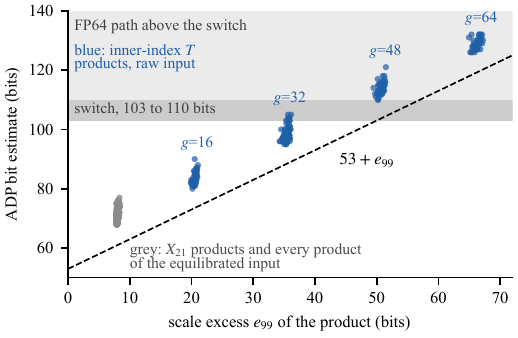}
\caption{ADP's estimate tracks $53 + e_{99}$, so it asks for the bits the inner-index scaling consumes; above the
switch the call runs in FP64.}
\label{fig:bits}
\end{figure}

\begin{figure}[t]
\centering
\includegraphics[width=\columnwidth]{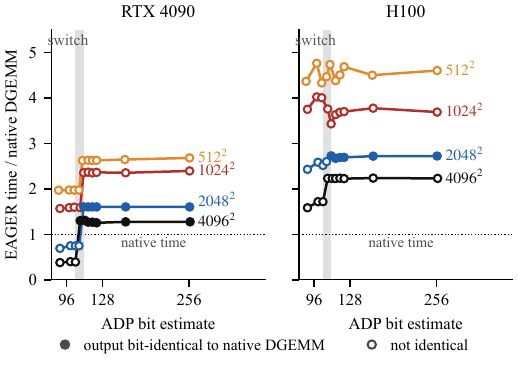}
\caption{Every output size of a single EAGER product switches path at the same reported estimate (shaded).}
\label{fig:dispatch}
\end{figure}

\paragraph{ADP gives $T$ the bits it consumes}
ADP's estimate tracks $53 + e_{99}$ (least-squares slope \V{r4090.fit.slope}, Fig.~\ref{fig:bits}). Single products
with a dialled-in inner-index span show what follows (Fig.~\ref{fig:dispatch}): every output size switches to an FP64
path at about the same estimate, $4096 \times 4096$ outputs between \V{r4090.thr.4096.emu} and
\V{r4090.thr.4096.cap} bits. That path is not free: it still computes the estimate, and on the H100 it is an FP64 GEMM
of the emulation library, so it takes longer than native DGEMM, most on small outputs (Appendix~\ref{app:switch}).
Each $T$ thus gets the bits it needs or runs in FP64, while the outer-index products keep emulating.

\subsection{Does it hold on other GPUs and on real matrices?}
\label{sec:breadth}

The same grid ran on all four GPUs at $n = 8192$, and every route on \V{l40s.realsum.n} real SPD matrices
(Table~\ref{tab:general} shows two of the three groups). Every route reaches the same error at $g = 64$ on every GPU,
to two digits, and loses bits at the same rate in $g$: the loss is a property of the arithmetic, not of one chip. It
persists at $n = 16384$ and over seeds. At $g = 64$, 14 and 16 moduli and FIXED 55 all reach about
\V{h100.fx.m16.g64}, so their error there is saturated; 20 moduli and FIXED 79 lower it by three to four orders, and
FIXED 111 to within a small factor of native (Appendix~\ref{app:tables}).

On the real matrices the raw fixed routes lose most on the census covariances (Table~\ref{tab:general}). ADP brings every matrix within a factor \V{l40s.realsum.guard.max} of native, and equilibrated 14
moduli every matrix but \V{l40s.realsum.eqm14.arg} within \V{l40s.realsum.eqm14.second}. On
\V{l40s.realsum.eqm14.arg} the dynamic range within rows survives diagonal scaling ($\kappa =
\V{rm.bcsstk23.kappa}$ after equilibration), equilibrated 14 moduli stay \V{l40s.realsum.eqm14.max} times native, and
the guarded and solve-based routes are the repairs to use (Appendix~\ref{app:tables}).

In every one of the \V{h100k.vs.nver} tested cuBLAS releases with FP64 emulation (cuBLAS 13.0 has none), the guarded
routes stay at native level and the fixed routes fail. cuSOLVER's own \texttt{potri}, set to FIXED 55 through its
emulation control, fails the same way, \V{h100k.csfx.b55.potri.g64} at $g = 64$, and in the pinned release
equilibration repairs it, so the shipped routine shows the loss under the same setting.

\subsection{What does the enclosure certify?}
\label{sec:encl}

Can a run be audited without a reference inverse? Theorem~\ref{thm:comp} is evaluated on each run's own operands, with
every product's error bounded by its graded ratios and every rounding enclosed. It gives a certified bound on the
error relative to $(LL^T)^{-1}$ for the stored $L$, and a run is certified when that bound is below 1. On every GPU and
size run it certifies at least \V{en.cert.n} of the \V{en.pairs} route--spread pairs (Table~\ref{tab:encl}); on the
H100 at $n = 4096$ these are all \V{en.acc.n} accurate pairs and \V{en.bad.cert.n} moderately failing ones, with no
bound only for the runs with the largest errors. The bound
exceeds the measured error by one to five orders of magnitude, more at larger $n$, as worst-case bounds do. The
enclosure is therefore a usable audit of a completed run, needing neither a reference inverse nor more precision than
the run itself (Appendix~\ref{app:grading}).

The evaluator itself was checked in exact arithmetic, against an independent second evaluator and by rebuilding
stored runs on a CPU (Appendix~\ref{app:grading}).

\begin{table}[t]
\caption{The enclosure certifies every accurate pair (route--spread pairs certified, of \V{en.pairs}; --: not run).}
\label{tab:encl}
\centering\footnotesize
\setlength{\tabcolsep}{4pt}
\begin{tabular*}{\columnwidth}{@{\extracolsep{\fill}}lccc@{}}
\toprule
GPU & $n = 2048$ & $n = 4096$ & $n = 8192$ \\
\midrule
H100 & \V{h100k.en2048.cert.n} & \V{en.cert.n} & -- \\
A100 & \V{a100.en2048.cert.n} & \V{a100.en4096.cert.n} & \V{a100.en8192.cert.n} \\
RTX 4090 & \V{x4090.en2048.cert.n} & \V{en4.cert.n} & \V{x4090.en8192.cert.n} \\
L40S & \V{l40s.en2048.cert.n} & \V{l40s.en4096.cert.n} & \V{l40s.en8192.cert.n} \\
\bottomrule
\end{tabular*}
\end{table}

\subsection{What does accuracy cost?}
\label{sec:cost}
\label{sec:full}

Is an accurate emulated inversion faster than native? Comparing complete device-resident routes that differ in one
factor answers this (Table~\ref{tab:fairm}, blocked LAUUM throughout; rows 2 and 3 multiply to row 4). Equilibration
is nearly free, which makes it the cheapest repair. Where FP64 is scarce, emulation pays: 16 moduli take
\V{x4090.d3x.eq2m16} of native's time on the RTX 4090, more than recovering the blocked algorithm's deficit against
cuSOLVER, so the emulated route takes \V{x4090.d3x.bestm16} of cuSOLVER's time. On the H100 the emulated route's gain
over cuSOLVER (\V{h100k.d3x.bestm16}) comes from the blocked algorithm, not from emulation: native blocked inversion
alone takes \V{h100k.d3x.eq2potri} of cuSOLVER's time, and native FP64 is fastest. Absolute times, transfers, workspace
and the cost of the other repairs are in Appendix~\ref{app:cost}.

\begin{table}[t]
\caption{Time ratios ($n = 8192$, $g = 64$): emulation wins where FP64 is scarce, the blocked algorithm where
it is plentiful.}
\label{tab:fairm}
\centering\scriptsize
\setlength{\tabcolsep}{2pt}
\begin{tabular*}{\columnwidth}{@{\extracolsep{\fill}}lcccc@{}}
\toprule
 & \multicolumn{2}{c}{FP64-rich} & \multicolumn{2}{c}{FP64-scarce} \\
time ratio & H100 & A100 & RTX 4090 & L40S \\
\midrule
native: equilibrated vs raw & \V{h100k.d3x.eq2cost} & \V{a100.d3x.eq2cost} & \V{x4090.d3x.eq2cost} & \V{l40s.d3x.eq2cost} \\
native, eq., vs cuSOLVER & \V{h100k.d3x.eq2potri} & \V{a100.d3x.eq2potri} & \V{x4090.d3x.eq2potri} & \V{l40s.d3x.eq2potri} \\
$q$=16 vs native, both eq. & \V{h100k.d3x.eq2m16} & \V{a100.d3x.eq2m16} & \V{x4090.d3x.eq2m16} & \V{l40s.d3x.eq2m16} \\
$q$=16, eq., vs cuSOLVER & \V{h100k.d3x.bestm16} & \V{a100.d3x.bestm16} & \V{x4090.d3x.bestm16} & \V{l40s.d3x.bestm16} \\
\bottomrule
\end{tabular*}
\end{table}

\subsection{Does the loss reach an application?}
\label{sec:app}

Does a loss in the inverse reach a quantity a user reads? Cokriging predicts one spatial variable from several
correlated ones and uses the inverse covariance for every prediction and its variance~\cite{wackernagel2003}. Five
variables of the California housing data~\cite{pace1997}, in the units of the file, give a covariance
$B_{\mathrm{cov}} \otimes (C + \tau I)$ whose diagonal spans many binades, with $C$ a squared-exponential kernel on
location. The nugget $\tau = 10^{-1}$ is the model, $10^{-3}$ a sensitivity case and $10^{-6}$ a stress case. The
measure is the RMS distance of held-out predictions from a certified long-double reference, as a fraction of the
predictor's own held-out RMSE (Appendix~\ref{app:cert}).

The weakest fixed routes move predictions \V{ckh.amp.min} to \V{ckh.amp.max} times as far from the reference as native
cuSOLVER does, over both sizes and all three nuggets (Table~\ref{tab:appm} shows $n = 4608$). ADP stays at native level, and equilibration stays
within a small factor of it. The loss is damped, because each prediction sees the inverse's error only through a
projection onto its covariance vector, and the realised share of the worst case falls as the inverse's error
grows (Appendix~\ref{app:cert}). At the model's nugget a certified
bound keeps the loss below \V{rb.t1.fixedcert} of the prediction error. An application with a smaller nugget, or one that reads the inverse's entries directly, as
kriging variances do, sees more of it.

\begin{table}[t]
\caption{Raw fixed routes move cokriging predictions; equilibration and ADP do not (H100, $n = 4608$).}
\label{tab:appm}
\centering\footnotesize
\setlength{\tabcolsep}{4pt}
\begin{tabular}{@{}lccc@{}}
\toprule
route & $\tau = 10^{-1}$ & $\tau = 10^{-3}$ & $\tau = 10^{-6}$ \\
\midrule
native, cuSOLVER & \V{ckh.lm.l3.t1.4608.cusolver.potri.dref} & \V{ckh.lm.l3.t3.4608.cusolver.potri.dref} & \V{ckh.lm.l3.t6.4608.cusolver.potri.dref} \\
EAGER, cuSOLVER & \V{ckh.lm.l3.t1.4608.cusolveremueager.potri.dref} & \V{ckh.lm.l3.t3.4608.cusolveremueager.potri.dref} & \V{ckh.lm.l3.t6.4608.cusolveremueager.potri.dref} \\
\midrule
$q$=14 & \V{ckh.lm.l3.t1.4608.blk.gemmul8m14.dref} & \V{ckh.lm.l3.t3.4608.blk.gemmul8m14.dref} & \V{ckh.lm.l3.t6.4608.blk.gemmul8m14.dref} \\
F55 & \V{ckh.lm.l3.t1.4608.blk.cublasemufixed55.dref} & \V{ckh.lm.l3.t3.4608.blk.cublasemufixed55.dref} & \V{ckh.lm.l3.t6.4608.blk.cublasemufixed55.dref} \\
\midrule
$q$=14, eq. & \V{ckh.lm.l3.t1.4608.blkeq.gemmul8m14.dref} & \V{ckh.lm.l3.t3.4608.blkeq.gemmul8m14.dref} & \V{ckh.lm.l3.t6.4608.blkeq.gemmul8m14.dref} \\
F55, eq. & \V{ckh.lm.l3.t1.4608.blkeq.cublasemufixed55.dref} & \V{ckh.lm.l3.t3.4608.blkeq.cublasemufixed55.dref} & \V{ckh.lm.l3.t6.4608.blkeq.cublasemufixed55.dref} \\
\midrule
certified bound, worst fixed & \V{rb.t1.fixedcert} & \V{rb.t3.fixedcert} & --- \\
\bottomrule
\end{tabular}
\end{table}

\section{Lessons and limits}
\label{sec:discussion}

\paragraph{For library builders}
Three rules follow. Check, for each product of a composite routine, which index its operand
scaling falls on, not only how large its error is. Prefer associations that keep the scaling on outer indices, such as
Method~1B, or equilibrate before emulated products. LAPACK's reference \texttt{dtrtri} already does so for an upper
factor: its upper branch on $R = R_MD$ is the transpose of Method~1B on $L$, and in the emulator model of
Section~\ref{sec:mechanism} it is bit-for-bit scale equivariant at every $g$ (\V{ac.up.inv} of \V{ac.up.cases}), with
errors at most \V{ac.up.errmax}, while its lower branch is Method~2C. Rely on a per-call guard such as ADP only where its estimate sees
the scale excess, as it does here. A guard that looked only at the outer scaling of each product, or a fixed bit
budget tuned on well-scaled inputs, would pass every per-product test and still fail here.

\paragraph{Limitations}
All four GPUs are NVIDIA's; Blackwell GPUs were not available to us, and cuBLAS's ADP is the only shipped emulation
guard we know of. Equilibration repairs FIXED 55 to native level only from cuBLAS 13.7 (13.8 on the H100). The closed libraries %
do not document Proposition~\ref{prop:solve}'s hypotheses; they are checked on outputs only. A precision selector built
on the enclosure is future work. The application uses one dataset and one covariance family.

\section{Conclusion}
\label{sec:conclusion}

Per-product accuracy grades do not compose for blocked triangular inversion, because its product form contracts the
diagonal scaling along the inner index of one product. Moving the scaling off that index, or giving the product the
bits it consumes, restores accuracy, and where FP64 is scarce the accurate emulated inversion still beats cuSOLVER.

\section*{Data availability}
An anonymized artifact holds the harness, the retained run records and the scripts behind each number, table and
figure, and \texttt{supplementary\_tables.pdf} with every table in full and the verification record:
\url{https://figshare.com/s/bb220965206030883d9e} (sha256 %
\texttt{c68e11946d3fe33f\allowbreak 51c8006806922fb0\allowbreak 5c084ea3b758ac35\allowbreak 17e47bb4f53fc76f}; %
the separate witness archives are listed in its README).

\section*{Use of Generative AI}
Claude Code (Anthropic; Claude Opus 5.5) assisted with writing experiment code and drafting and revising the
text; the authors reviewed and revised all of it, and every number in the paper is generated from the result
files by a script. ChatGPT (OpenAI; GPT-5.6 Sol) was used for critique of drafts. %


\appendices
\section{Setup and evidence}
\label{app:setup}
\label{app:audit}
\label{app:evid}

\paragraph{Platforms and libraries}
All four GPUs run the same pinned Python wheels (nvidia-cublas 13.8.1.7, nvidia-cusolver 12.3.5.3, cudart %
13.4.92); the H100 also runs a CUDA 13.2 module's libraries. Drivers, clocks, power limits and NumPy and SciPy %
versions are in the supplement (\texttt{supplementary\_tables.pdf} in the artifact). GEMMul8 v3.5.4 uses its INT8 back
end with accurate scaling for $q = 14, 16, 20$ moduli. Host LAPACK is OpenBLAS from the PyPI wheels, which a harness
test checks to be bit-for-bit scale equivariant on every job.

\paragraph{Routes}
cuSOLVER \texttt{potri} runs on LAPACK's factor natively, under ADP with DEFAULT and with EAGER, and EAGER on the
equilibrated factor. A blocked TRTRI and LAUUM route every GEMM through one back end. Recursion halves down to
256-wide leaves. LAUUM is either one $X^TX$ GEMM, whose operands are both triangular, or blocked by the same
recursion, which leaves no product with zero triangles on both operands; every route uses the single GEMM unless
marked ``blocked LAUUM''. Hybrid routes compute every $T$, or every $X_{21}$, by native DGEMM. The solve-based routes
use the association of Proposition~\ref{prop:solve}. Both implementations use the lower triangle, as do the cuSOLVER
routes, and perform each update and negation as a separate operation (NumPy; \path{cublasDgeam}).

\paragraph{Inputs}
The scaled class is $DMD$ with $M = W^TW/n + I$ for $W \in \mathbb{R}^{n \times n}$ with independent standard normal
entries and $k_i$ uniform on $\{0,\dots,g\}$. Seed 0 covers the grid of Table~\ref{tab:setup} at $n \in \{2048, 4096,
8192\}$; seeds 1 to 4 repeat $g \in \{0, 16, 32, 64\}$ at $n = 4096$ and $g = 64$ at $n = 8192$; the breadth runs add
$n = 16384$. Ordering controls reorder the same exponents. The mixed-units class scales the rows and columns of $M$ by
unit factors between $10^{-6}$ and $10^{6}$, not powers of two; its diagonal spans $2^{\V{um.diag.log2}}$, about $g =
\V{um.g}$.

\paragraph{Reference and metrics}
The reference $D^{-1}X_MD^{-1}$ is scale-consistent with the input. Against inverses computed with \V{ref.dps} decimal
digits for $n \le \V{ref.nmax}$ and $g \in \{0, 32, 64\}$, it is within \V{ref.ref.norm.max} normwise and
\V{ref.ref.cw.max} componentwise, as is LAPACK's own inverse of $A$. Elsewhere the reference is LAPACK's inverse of $A$.
$\widehat X$ is the symmetric computed inverse. A reference-free check is the scaled inverse residual: with $\widehat Y
= D\widehat XD$, exact for power-of-two $D$, it is $\|I - \widehat YM\|_F/(\|M\|_F\|\widehat Y\|_F)$, formed by native
DGEMM on the device for every full route. 

\paragraph{Timing}
Full routes run on the device from the uploaded matrix to the inverse: cuSOLVER's potrf (native), any power-of-two
equilibration by \path{cublasDdgmm}, the inversion and any rescaling. Each route is also timed with the upload of $A$
and the download of the inverse from pageable host memory inside the clock. Device memory is sampled while each route
runs once in a fresh process. The H100 ran \V{d3.h100.sessions} independent sessions (separate jobs, $n = 8192$, $g \in
\{0, 64\}$), and an earlier session gives a further one for the routes it shares;\ifnum\V{d3.r4090.pending}=1{} the RTX
4090 figures come from that earlier session alone.\else{} the RTX 4090 ran \V{d3.r4090.sessions} sessions at $g = 64$
only.\fi{} The bit-count pointer is not installed in timing runs.

\begin{table}[t]
\caption{Evidence coverage. Routes: timed routes whose strongest link it is (H100 / RTX 4090).}
\label{tab:cover}
\centering\scriptsize
\setlength{\tabcolsep}{2pt}
\begin{tabular}{@{}>{\raggedright\arraybackslash}p{0.21\columnwidth}>{\raggedright\arraybackslash}p{0.6\columnwidth}>{\centering\arraybackslash}p{0.15\columnwidth}@{}} %
\toprule
evidence & executions and coverage & routes \\
\midrule
product grades & harness, $n = \V{en.n}$, H100: every output entry of every TRTRI product; LAUUM on one sampled block
per run & \\[1.5pt]
full-output grades and enclosure & harness, $n = \V{en.n}$, both GPUs, \V{en.pairs} route--spread pairs and
\V{en.hyb.n} hybrids: every output entry of every TRTRI and LAUUM product; certificate from those grades &
\V{em.h100.n.harness} / \V{em.r4090.n.harness} \\[1.5pt]
other inverse errors & harness, all GPUs and sizes: $\epsilon_X$ only & \\[1.5pt]
device audit & device-resident, RTX 4090: TRTRI products of \V{au.routes} routes at $n = \V{en.n}$, $g \in \{0, 64\}$
& \V{em.h100.n.A4} / \V{em.r4090.n.A4} \\[1.5pt]
captured timed execution & device-resident, $n = 8192$, $g = 64$: every product of the timed execution copied back
and graded; decisive routes also traced call by call & \V{em.h100.n.captured} / \V{em.r4090.n.captured} \\[1.5pt]
timing & device-resident, $n = 8192$, all GPUs: time, $\epsilon_X$ and residual; products not graded &
\V{em.h100.n.timing} / \V{em.r4090.n.timing} \\
\bottomrule
\end{tabular}
\end{table}

\paragraph{The implementation audit}
Table~\ref{tab:cover} states what each kind of evidence covers. On the RTX 4090 ($n = 4096$, $g \in \{0, 64\}$) every
TRTRI product of \V{au.routes} device-resident routes was copied back after its call, which leaves $L^{-1}$
bit-identical, and graded with the harness's code: the largest $r_B^{\mathrm{up}}/k$ is \V{au.rb.dev} on the device
(harness: \V{au.rb.host}), and at $g = 64$ the largest $r_A/k$ on $T$ for 14 moduli is \V{au.ra64.dev} (harness:
\V{au.ra64.host}). The TRSM leaves run natively: on handles carrying DEFAULT, EAGER or FIXED 55 a leaf TRSM issues its
internal products with the native compute type (\V{tr.trsm.native} of \V{tr.trsm.n} configurations) and launches
exactly native's kernels (\V{tr.kern.same} of \V{tr.kern.n}). The operations Proposition~\ref{prop:solve} assumes were
tested through their observable input/output equivariance, bit for bit, at $n = 4096$, $g = 64$. Hypothesis (i) held on every product of
the solve-based route for every library (\V{eqv.outer} of \V{eqv.outern}), while inner scaling changed every emulated
product (\V{eqv.inner} of \V{eqv.innern} unchanged).

\ifnum\V{ap.h100.done}=1\ifnum\V{ap.r4090.done}=1{}A paired audit ties a failing timed route to its repair on each GPU
($n = 8192$, $g = 64$): FIXED 55 and equilibrated FIXED 55 (one-GEMM LAUUM) on the H100, 14 moduli and equilibrated 14
moduli on the RTX 4090, with every product of the timed execution captured and graded on its full output. LAUUM's
product meets $r_A/k \le \V{ap.l.ra.max}$ on all four routes, and the failing routes' $T$ products miss grade A by up
to \V{ap.h100.fail.t.ra} (H100) and \V{ap.r4090.fail.t.ra} (RTX 4090), with inverse errors \V{ap.h100.fail.err} and
\V{ap.r4090.fail.err} against \V{ap.h100.rep.err} and \V{ap.r4090.rep.err} repaired: the loss sits in the $T$ products
of the timed execution, as in the harness.\fi\fi{}

\ifnum\V{br30.r4090.done}=1{}The four decisive routes on each GPU were rerun in separate jobs with every kernel of
every GEMM call traced: in \V{br30.r4090.match} of \V{br30.r4090.calls} traced calls on the RTX 4090
\ifnum\V{br30.h100.done}=1{}and \V{br30.h100.match} of \V{br30.h100.calls} on the H100 \fi{}the kernel sequence equals
that of the captured execution, and every returned inverse of a route is equal across untraced, timed, traced and
captured runs.\fi{} The supplement gives the full audit and the per-route evidence matrix.

\paragraph{A leftover setting}
cuBLAS keeps the maximum mantissa bit count of a FIXED configuration on the handle after the setting returns to ADP,
and the harness, which uses one handle per process, did not restore it. A dynamic route that followed a FIXED route in
the same process therefore ran under a cap of 55 or 79 bits: EAGER at $n = \V{en.n}$, $g = 32$ on the H100 has an %
enclosure of \V{lk.fresh.rel} in a fresh process and \V{lk.capped.rel} after FIXED 55. The harness now restores and
asserts the default at every emulated call and records the effective cap per route. Every set from which a printed
number of a dynamic route comes was rerun, and the artifact's provenance check fails if a printed number comes from a
capped row. The reruns brought EAGER's grades at $g = 32$ and 64 to their level at $g = 0$ and moved errors in the
last printed digit at most; the superseded records and the rerun list are in the artifact.

\section{Theory in full}
\label{app:full}

\paragraph{Proposition~\ref{prop:cex}: what the bounds admit}
With this paper's $f(k) = k$ the bound admits $|\delta|$ up to $2^{g-52}$; at $g = 64$ already $\delta = \V{ce.delta}$ gives \V{ce.err}. Grade A admits at most %
$f_A(2)\,u\,p$ with $p = 2$ on the same entry, and $(|L^{-1}||L||L^{-1}|)_{31}$, which bounds classical triangular
inversion~\cite{ducroz1992,higham2002}, is \V{ce.cw} for every $g$; a script checks all of this in exact rationals for
$g$ from 0 to 64. The proposition concerns what the bound permits, not any library, and the error of $L^{-1}$; the
measurements below concern $Y = X^TX$, which the perturbation reaches at the same order.

\paragraph{Proof of Theorem~\ref{thm:comp}}
The remainder of (a) beyond first order is $R = -\sum_s\Lambda_s\Delta_{22}L_{21}\Delta_{11}\Gamma_s$. Part (c)
claims nothing about how the frame quantities vary with $D$ for a given implementation.

\begin{IEEEproof}
(a) $X_{21} = -X_{22}L_{21}X_{11}$ and $\widehat X_{21} = -\widehat X_{22}L_{21}\widehat X_{11} + \Phi_s$; substituting
$X_{ii} = \widehat X_{ii} - \Delta_{ii}$ and $\widehat X_{22}L_{21} = \widehat T - E_s$ gives \eqref{eq:rec}. Trivially
$\Delta = X(L\Delta L)X$, and by \eqref{eq:rec} and $L_\nu X_\nu = I$ on each node the block $(\rho_2, \rho_1)$ of
$L\Delta L$ is $L_{22}\Psi_sL_{11}$ (the identity below). (b) Take absolute values in \eqref{eq:rec}, by induction
from the leaves, and in $\widehat Y - Y = \widehat X^T\Delta + \Delta^T\widehat X - \Delta^T\Delta + \mathrm{sym}(G) +
\varrho$, where $|\mathrm{sym}(G)| \le \mathrm{sym}(\mathcal G)$ (not $\mathcal G$, unless it is symmetric) and the
rounding $\varrho$ of the symmetrization satisfies $|\varrho| \le 2u|\widehat Y| + 2^{-1074}$ entrywise: the addition %
errs by at most a relative $u$ and is exact if its result is subnormal, and the halving errs by at most $2^{-1075}$. %
(c) Substitute $L = DL_M$ and $\widehat X = \widehat X'D^{-1}$: every $D$ cancels, for instance $B_{22}|L_{21}\widehat
X_{11}|D_1 = B'_{22}|L_{M,21}\widehat X'_{11}|$ and $|\widehat X_{22}||L_{21}| = |\widehat X'_{22}||L_{M,21}|$. Under
grade B on $T$ the allowance is $c\,u\,r_is_j$ with $r_i = \max_k|\hat x'_{ik}|/d_k$ and $s_j = \max_kd_k|l_{M,kj}|$
over $k \in \rho_2$, at most $\rho_s$ times its value at $D = I$.
\end{IEEEproof}

\eqref{eq:rec} is the forward-error counterpart of Du Croz and Higham's residual analysis of Method~2B, kept exact and
with arbitrary product errors. Appendix~\ref{app:grading} evaluates a sharper form of (b) that keeps $\Lambda_\nu$ and
$\Gamma_\nu$. Three limits of (c) matter. What is free of $D$ is the enclosure of the frame error $D|\widehat Y - Y|D$ as a
function of the frame's computed quantities; these lie within $B'$ and $\mathcal E_s$ of the $D$-free $L_M^{-1}$ and
$L_{M,22}^{-1}L_{M,21}$, so the enclosure is independent of $D$ to first order in the allowances. The unscaled
normwise $\epsilon_X$ weights the entries by $D^{-1}$ and has no such property. And the condition is sufficient, not
necessary; Proposition~\ref{prop:cex} attains the $\rho_s$ growth, so under grade B alone $D$ cannot be removed.

\emph{The identity.} Fix a node of the recursion with contiguous index set $\nu = \rho_1 \cup \rho_2$. $L$ and
$\Delta$ are lower triangular, so the block $(\rho_2, \rho_1)$ of $L\Delta L$ involves only indices in $\nu$ and equals
$L_{21}\Delta_{11}L_{11} + L_{22}\Delta_{21}L_{11} + L_{22}\Delta_{22}L_{21} = L_{22}(\Delta_{21} +
X_{22}L_{21}\Delta_{11} + \Delta_{22}L_{21}X_{11})L_{11}$. Expanding $X_{22}L_{21}X_{11}$ with $X_{ii} = \widehat X_{ii}
- \Delta_{ii}$ gives $\Delta_{21} = \Phi_s - \widehat X_{22}L_{21}\Delta_{11} - \Delta_{22}L_{21}\widehat X_{11} +
\Delta_{22}L_{21}\Delta_{11}$, which is \eqref{eq:rec} and, with $\widehat X = X + \Delta$ once more, $\Delta_{21} +
X_{22}L_{21}\Delta_{11} + \Delta_{22}L_{21}X_{11} = \Phi_s - \Delta_{22}L_{21}\Delta_{11} = \Psi_s$. Every strictly
lower entry lies in the block $(\rho_2, \rho_1)$ of exactly one split and every other entry of the lower triangle in
one leaf, where the block of $L\Delta L$ is $L_\ell\Delta_\ell L_\ell$. Hence $\Delta = X(L\Delta L)X =
\sum_\nu X_{:,\rho_2}L_{22}\Psi_\nu L_{11}X_{\rho_1,:}$. Both identities are polynomial; a script checks
\eqref{eq:rec} symbolically and both as exact rational equalities on \V{en.self.cases} random cases.

\emph{The bound that is evaluated.} $\widehat Y - Y = X^T\Delta + \Delta^TX + \Delta^T\Delta + \mathrm{sym}(G)$ and
$X^T\Delta = \sum_\nu (X^T\Lambda_\nu)\Psi_\nu\Gamma_\nu$. With $X = \widehat X - \Delta$ and $|\Delta| \le B$
from \eqref{eq:B},
\begin{align*}
|X^T\Lambda_\nu| &\le |\widehat X^T\widehat X_{:,\rho_2}L_{22}| + |\widehat X|^TB_{:,\rho_2}|L_{22}|\\
 &\quad + B^T\bigl(|\widehat X_{:,\rho_2}L_{22}| + B_{:,\rho_2}|L_{22}|\bigr) =: \mathcal Y_\nu,\\
|\Gamma_\nu| &\le |L_{11}\widehat X_{\rho_1,:}| + |L_{11}|B_{\rho_1,:} =: \bar\Gamma_\nu,\\
|\Psi_s| &\le \mathcal E_s|\widehat X_{11}| + \mathcal F_s + B_{22}|L_{21}|B_{11} =: \bar\Psi_s,
\end{align*}
and $\bar\Psi_\ell = B_\ell$, so that
\[
\begin{aligned}
|\widehat Y - Y| \le{}& \textstyle\sum_\nu \bigl(\mathcal Y_\nu\bar\Psi_\nu\bar\Gamma_\nu + (\cdot)^T\bigr) + B^TB\\
&+ \mathrm{sym}(\mathcal G) + 2u|\widehat Y| + 2^{-1074}\mathbf 1\mathbf 1^T, %
\end{aligned}
\]
the last two terms covering the rounding of the symmetrization, as in Theorem~\ref{thm:comp}(b). Without the absolute %
term the bound is false: for $\widehat X = X = I_2$, $B = 0$ and an output error $G$ with off-diagonal entries
$2\alpha$ and $\alpha$, $\alpha = 2^{-1074}$, the halving rounds $3\alpha/2$ to $2\alpha$, which exceeds %
$\mathrm{sym}(\mathcal G)_{12} + 2u|\widehat Y_{12}|$; the self-tests check this example. The evaluator enters the
absolute term raised to its floor $\tau = 2^{-500}$, a larger and therefore valid allowance. The symmetrization of %
the allowance matters: for $\widehat X = I_2$ and $G = e_1e_2^T$, $\mathrm{sym}(G)$ has the entry $1/2$ in position %
$(2, 1)$, where the envelope $|G|$ is zero. The allowance used here, $\min(r_A^{\mathrm{up}}u|\widehat X|^T|\widehat X|,
r_B^{\mathrm{up}}u\,cc^T)$ with $c$ the column maxima of $|\widehat X|$ and one pair of ratios for the whole product,
is symmetric in exact arithmetic; the code nevertheless replaces whatever allowance it is given by
$\max(\mathcal G, \mathcal G^T) \ge \mathrm{sym}(\mathcal G)$ and asserts the result symmetric (largest relative
asymmetry of the computed allowance before that step: \V{en.asym}). The terms of first order in the allowances keep
$X^T\Lambda_\nu$ and $\Gamma_\nu$ formed before the absolute value, with the run's $\widehat X$ in place of the exact
inverse; $B$ enters at second order only.

\emph{Leaves.} With $N \ge |\widehat X_\ell L_\ell - I|$ and $\|N\|_\infty = \nu_\ell < 1$, $\Delta_\ell = (\widehat
X_\ell L_\ell - I)(\widehat X_\ell - \Delta_\ell)$ gives $|\Delta_\ell| \le (I - N)^{-1}N|\widehat X_\ell| \le N|\widehat
X_\ell| + \tfrac{\nu_\ell}{1 - \nu_\ell}(Ne)(e^T|\widehat X_\ell|) =: B_\ell$. The residual is formed in long double and
$N$ includes its rounding, $\gamma^{ld}_{n_b+1}(|\widehat X_\ell||L_\ell| + I)$; in the runs below $\nu_\ell \le
\V{en.leaf.nu}$.

\paragraph{Proposition~\ref{prop:solve} in full}
In Method~1B each update forms $P = L_bZ_1$ by GEMM with $\beta = 0$ and then subtracts, $W_2 - P$, as a separate
operation; the leaves are native TRSM of width $n_b$.
The full statement: Run the solve-based recursion on $L_M$ and on $DL_M$, $D$ a power-of-two diagonal, with the same partition and the
same sequence of operations, in one binary floating-point format and rounding mode~\cite{ieee754} (binary64, round to
nearest, in the tested libraries) with no trap or flag altering control flow, the entries of $D$ and $DL_M$ being
exact. Assume in both runs that (i) every product is
computed by an outer-equivariant $G$; (ii) each update and each negation is one IEEE operation per entry, neither
fused or contracted with the product it follows nor reassociated; (iii) the leaves are equivariant: the inversion
returns $\widehat X(L_a)D_a^{-1}$ for $D_aL_a$, and the solve returns $\widehat Z(L_a, W)C$ for $(D_aL_a, D_aWC)$,
$D_a$, $C$ power-of-two diagonal; (iv) every routine is deterministic and takes the same computational path in both
runs, with the same order of operations inside every reduction, and no operation, intermediate ones and the symmetrization included, overflows, underflows, returns a
subnormal or raises another exception. Then $\widehat X(DL_M) = \widehat X(L_M)D^{-1}$ bit for bit, and LAUUM on $G$
with symmetrization returns $D^{-1}\widehat Y(L_M)D^{-1}$. If also (v) the computed Cholesky factor of $DMD$ is $D$
times that of $M$, the computed inverse of $DMD$ is $D^{-1}$ times that of $M$ times $D^{-1}$, bit for bit ($\pm0$ equal).

\begin{IEEEproof}
By induction on the recursion; the leaves are (iii). If $\widehat X_{ii} = \widehat X_{M,ii}D_i^{-1}$, then $W =
G(D_2L_{M,21}, \widehat X_{M,11}D_1^{-1}) = D_2W_MD_1^{-1}$ by (i). In the solve, a leaf maps $(D_aL_{M,a},
D_aW_{M,a}D_1^{-1})$ to $\widehat Z_{M,a}D_1^{-1}$ by (iii); an update forms $G(D_bL_{M,b}, \widehat Z_{M,a}D_1^{-1}) =
D_bP_MD_1^{-1}$, and a correctly rounded subtraction or negation of operands sharing power-of-two factors returns the
factors times the unscaled result under (ii) and (iv), so $\widehat X_{21} = \widehat X_{M,21}D_1^{-1}$. LAUUM's
$G(\widehat X^T, \widehat X)$ carries $D^{-1}$ on outer indices only, and $(Y + Y^T)/2$ commutes with the scaling. (v)
closes the pipeline.
\end{IEEEproof}

(i), (iii), (iv) and (v) are properties of library routines that no documentation guarantees; Appendix~\ref{app:setup}
checks them on outputs, bit for bit, which is evidence for the tested versions, configurations and inputs and not a
guarantee for others. Under equivariance $\widehat Y(DMD) -
(DMD)^{-1} = D^{-1}(\widehat Y(M) - M^{-1})D^{-1}$, so every entry's relative error, and the error measured in $M$'s
frame, equal those of $M$'s inversion, whatever $D$; the normwise $\epsilon_X$ need not be constant: for $M = I$, an error $\eta e_1e_1^T$ in $M$'s inverse and $D =
\mathrm{diag}(1, 2^g, 2^g)$, it rises from $|\eta|/\sqrt 3$ to nearly $|\eta|$. The product form has no such property,
because outer equivariance says nothing about a scaling shared along the contracted index. In the emulator run of
Section~\ref{sec:mechanism} the solve with $L_{11}$ is native, and with native products both Method~2C and the
product form are bit-for-bit scale equivariant and accurate (\V{ac.nat.inv} of \V{ac.nat.cases} runs).

\paragraph{The triangle}
LAPACK here means the reference \texttt{dtrtri}, characterized by its source-level
association; nothing is claimed about vendor implementations, which were not inspected. For both values of
\texttt{uplo}, the reference \texttt{dtrtri} multiplies the already inverted
block into the triangle's off-diagonal block (\texttt{DTRMM}) and then solves with the diagonal block
(\texttt{DTRSM}), contracting the triangle's row index; it never calls \texttt{DGEMM}, so it is exposed only where
those two routines are themselves emulated. The lower factor $L = DL_M$ carries $D$ on its rows, the upper factor $R = R_MD$ on its
columns. Section~\ref{sec:discussion} states the upper branch's result in the emulator model of
Section~\ref{sec:mechanism}, with the triangular product emulated and the solve and the leaves native; it is a
statement about the reference association in that model. It does not make
\texttt{uplo = 'U'} a repair: the contracted index also depends on how a library forms its triangular products and
solves and how it inverts its diagonal blocks, and a left triangular solve against the identity, for instance,
contracts the triangle's column index, which is the scaled one for $R$. Compiled LAPACK and cuSOLVER with the upper
triangle were not measured in this paper.

\paragraph{A first-order counterfactual}
Before the enclosure, the first-order terms were evaluated with LAPACK's binary64
inverse in place of the exact one and the H100's graded ratios (LAUUM's from its sampled block). That value is not a
bound. It exceeds the measured $\epsilon_X$ in \V{cb.m.dom} of \V{cb.m.pairs} pairs, by \V{cb.m.over.lo} to
\V{cb.m.over.hi} times. Assigning native DGEMM's graded ratios to the products that a replacement plan computes natively, without
rerunning, it is, over \V{cb.r.n} plans, at most \V{cb.r.nat.max} for every plan
measured at native level and at least \V{cb.r.oth.min} for every other; for 14 moduli at $g = 64$ it falls from
\V{cb.r.blkgemmul8m14.none} to \V{cb.r.blkgemmul8m14.innerall} with every $T$ native and stays at
\V{cb.r.blkgemmul8m14.outerall} with every $X_{21}$ native. This orders replacement plans without running them; it
certifies nothing.

\section{Grading and certification}
\label{app:grading}
\label{app:thm}

\paragraph{Grading}
Table~\ref{tab:grade} grades the blocked inversion's products at $n = \V{gr.n}$: all \V{gr.trtri.products} TRTRI
products over all routes and $g$, on every output entry, and LAUUM's $n \times n$ product on one randomly sampled $512
\times 512$ output block per run (\V{gr.lauum.products} blocks). The reference is a sequential x87 long-double product,
whose error the next paragraph bounds by $\gamma^{ld}_kp_{ij}$. The grading therefore bounds the true ratio entry by entry, $r_B^{\mathrm{up}} = \max_{ij}\,(|\hat c_{ij} -
c^{ld}_{ij}| + \gamma^{ld}_k p_{ij})/(ur_is_j)$, and $r_A$ from above and below; an independent outward-rounded script
recomputes these bounds to within \V{gv.reldiff}. No classification depends on the reference's own error (\V{gr.bflip}
grade-B and \V{gr.aflip} grade-A verdicts change), nor on doubling the allowance to $\gamma^{ld}_{2k}$ (\V{gr.bflip2}
and \V{gr.aflip2}). Maximized over all ten routes, all three $g$ and every graded entry, $r_B^{\mathrm{up}}/k$ is at
most \V{gr.up.t} on the \V{gr.up.t.n} products $T$, \V{gr.up.x} on the \V{gr.up.x.n} products $X_{21}$ and
\V{gr.up.lauum} on the \V{gr.up.lauum.n} sampled LAUUM blocks. A second run grades every output entry of every
product, LAUUM's $n \times n$ product included, with the doubled allowance: $r_B^{\mathrm{up}}/k$ is at most
\V{en.up.t}, \V{en.up.x} and \V{en.up.lauum} on the H100 and \V{en4.up.t}, \V{en4.up.x} and \V{en4.up.lauum} on the
RTX 4090 (\V{en.bfail} and \V{en4.bfail} products fail). The ratios are rounded outward at every step in long double,
the conversion to binary64 included, so a stored ratio bounds the true one. The ratios are taken over every graded
output entry, zeros included; an entry whose right-hand side is zero counts as 0 if its error is zero and fails
otherwise.

\paragraph{The reference's rounding model}
The reference is NumPy's long-double matrix product, which has no BLAS
behind it: a compiled loop that, per entry, adds the $k$ products one by one to a sum that starts at zero. On x86-64
the type is the 80-bit x87 format, whose multiply and add each round once to a 64-bit significand, and x87 has no
fused multiply-add, so no contraction is possible. A term's relative perturbation is therefore a product of at most
$k$ factors $1 + \delta$, $|\delta| \le u_{ld}$ (one for its product, at most $k - 1$ for the additions after it), and
the error is at most $\gamma^{ld}_kp_{ij}$~\cite[Section~3.1]{higham2002}, negligible in $r_A/k$ units (at most
\V{gr.allow.a}) but not, for generic operands, in $r_B/k$ units at large $k$: the standard bound for rounded products
followed by rounded additions is already $\gamma_k$, in any order of summation, and a fused accumulation has the same
bound. A script compares the library
product bit for bit with that explicit sequence in long-double scalars and with a correctly rounded fused
accumulation built in rationals: \V{ld.seq} of \V{ld.entries} entries equal the former, and the two models differ on
\V{ld.differ} of them; the largest error is \V{ld.worst} of $\gamma^{ld}_kp_{ij}$. The model depends on the platform
and the build: the type must be the x87 80-bit format, the loop must not be vectorized, reassociated or evaluated in a
wider or narrower precision, and the x87 precision control must be at its 64-bit default. The grading code stops
unless the type has a 64-bit significand, and each full-output job first runs the bitwise comparison on its own host
and build and stops, grading nothing, if a single entry differs (\V{ld.jobs} jobs on \V{ld.hosts} hosts,
\V{ld.jobs.seq} of \V{ld.jobs.entries} entries equal). On a platform where long double is binary64 or binary128,
grading therefore fails closed.

\paragraph{The componentwise error}
The componentwise error is
$\max |\hat x_{ij} - x_{ij}|/|x_{ij}|$ over the entries whose scaled value $|d_id_jx_{ij}|$ exceeds $10^{-8}$ of the %
largest, so that no entry below roundoff in its own scale, and no zero, enters. On a transparent CPU Ozaki-I emulator (7-bit slices, exact integer
slice products, $n = 2048$) the largest TRTRI $e_{99}$ rises from \V{dc.g0.e99} at $g = 0$ to \V{dc.g64.e99} at $g = 64$,
the scaling term carrying the rise.

\paragraph{The certificate}
Each product's error is bounded entrywise by $\min(r_A^{\mathrm{up}}u|A||B|, r_B^{\mathrm{up}}u\,rs^T)$ with its own
graded ratios, so the bound of Appendix~\ref{app:full} can be evaluated on a run's own operands: the factor $L$ the
route was given and its computed leaves, products and inverse, with the leaves' errors enclosed from their residuals.
The evaluator runs in binary64, round to nearest. It returns a certificate only if every operation was evaluated under
the assumptions of the error bound used for it; otherwise it refuses and returns no number.
\begin{itemize}
\item \emph{Invariant.} Every nonnegative matrix it forms is checked and normalized before further use: it refuses on
  an entry that is NaN, infinite, negative or above $2^{500}$, and raises nonzero entries below $\tau = 2^{-500}$ to %
  $\tau$, which is sound because every bound is monotone in each nonnegative operand. Signed inputs must lie in
  $[2^{-480}, 2^{480}]$, so no multiplication inside a matrix product underflows and nothing overflows. %
\item \emph{Enlarged operations.} Every product and sum of nonnegative operands is multiplied by a factor that covers
  its rounding for any summation order, with or without fused operations~\cite{higham2002}; an absolute product
  $|AB|$ is enclosed by $|\mathrm{fl}(AB)|$ plus $\gamma_k|A||B|$; the allowance $\min(r_A^{\mathrm{up}}u|A||B|,
  r_B^{\mathrm{up}}u\,rs^T)$ carries a margin of $1 + 8u$. %
\item \emph{Steps.} (1) leaf bounds from long-double residuals, refused unless $\nu_\ell < 1/2$; (2) splits, children
  first, by \eqref{eq:B}; (3) the sum over splits and leaves of the sharper form above; (4) LAUUM's allowance,
  symmetrized, with the rounding of the symmetrization; (5) the two Frobenius norms, $\beta \ge \|W\mathcal BW\|_F$ and
  $\eta \le \|W\widehat YW\|_F$ in the route's output scaling $W$, each formed as an exact sum of exact squares with a
  directed square root; (6) $\beta/(\eta - \beta)$ in exact rationals.
\end{itemize}
The status is \emph{certified} when that quotient is below 1, \emph{no relative bound} when $\eta \le \beta$, and
\emph{refused} on any failed check. By the invariant every enlargement is applied to an operation that satisfies the
standard model, so by induction every normalized matrix dominates the exact quantity it stands for; steps 5 and 6
add no rounding. The claim rests on IEEE binary64 arithmetic in the BLAS and NumPy with conventional products, on the
x87 model of the grading reference, and on the route forming its products from the operands the harness passes it; the
residual argument of step 1 is the standard one of verification methods~\cite{rump2010}. The supplement gives the
evaluator operation by operation, a reference procedure and test vectors.

\paragraph{How the evaluator was checked}
In exact rationals, on \V{en.self.cases} random triangular systems with diagonal spreads up to $2^{40}$ and product %
errors up to \V{en.self.magmax} of the grade-B right-hand side, the evaluated bounds dominate $|\Delta|$ and
$|\widehat Y - Y|$ entrywise in every case (\V{en.self.ok} of \V{en.self.cases}). On a CPU Ozaki-I emulator (\V{en.emu.cases} cases) the bounds
dominate in every case and exceed the actual error by \V{en.emu.over.lo} to \V{en.emu.over.hi} times. Adversarial tests
inject a NaN, an infinity or a negative value into each of the \V{adv.stages} intermediates (\V{adv.inj} injections,
every one refused), and both evaluators pass every one of \V{vec.n} test vectors with exact expected outcomes
(\V{vec.problems} misses). An independent second evaluator, sharing no code, agrees in status and bound on
\V{en.sec.ok} of \V{en.sec.n} pairs on the H100 \ifnum\V{en4.v2}=1{}and \V{en4.sec.ok} of \V{en4.sec.n} on the RTX 4090
\fi{}(largest relative difference \V{en.sec.maxdiff}). At $n = 64$
the operands of \V{e2.h100.n} runs\ifnum\V{e2.both}=1{} on the H100 and \V{e2.r4090.n} on the RTX 4090\fi{} were
stored and every bound recomputed in exact integer arithmetic: the exact bound dominates the true error entrywise,
and the evaluator's status equals the exact one (\V{e2.h100.ok}\ifnum\V{e2.both}=1{} and \V{e2.r4090.ok}\fi{} runs
without a discrepancy). Five $n = 4096$ runs and six drawn by a preselected rule were rerun with every GPU output stored
and rebuilt on a CPU from those outputs alone: \V{aud.ident} of \V{aud.drawn} drawn runs and all five witnesses
reproduced their status and bound bit for bit\ifnum\V{wind.done}=1{}, and on an independent x86-64 host
\V{wind.status} of \V{wind.n} witness statuses reproduced\fi. A preflight on the evaluating host (\V{pf.checks}
checks) gates every certificate. The archives, their sha256 values and the selection rules are in the artifact's
README and the supplement.

\paragraph{Results}
Table~\ref{tab:grade} lists the enclosure for every H100 pair. No pair was refused and none has a finite
relative bound of 1 or more: the \V{en.vac.n} pairs without a certificate all fail the denominator test,
with $\beta/\eta$ from \V{en.vac.over.lo} to \V{en.vac.over.hi} on the H100. In $M$'s frame the native route's enclosure
is \V{en.frame.native.g0}, \V{en.frame.native.g32} and \V{en.frame.native.g64} at $g = 0$, 32 and 64. The first-order part alone, with no leaves,
no $B$ and no rounding terms, is within \V{en.rem.max}\% of the enclosure wherever the enclosure is below
\V{en.rem.thr}; the leaves contribute at most \V{en.leaf.max}. Above that threshold the matrix $B$, which takes
absolute values at every level of the recursion, makes the second-order terms dominate: FIXED 79 at $g = 64$ is
certified at \V{en.rem.big.max}\% above its first-order part, and 20 moduli at $g = 64$, whose first-order part is %
below 1, has no relative bound.

\section{Results in full}
\label{app:tables}
\label{app:more}
\label{app:controls}

\paragraph{Grades and enclosure}
Table~\ref{tab:grade} gives all ten routes at three spreads; grading every entry of LAUUM too gives a B-type ratio of
at most \V{en.up.all}; native and ADP products meet $r_A/k \le \V{gr.native.amax}$ at every $g$. Over the
\V{gr.pred.count} route--spread pairs whose worst A-type ratio over the TRTRI products exceeds 100, the inverse's error is %
\V{gr.pred.lo} to \V{gr.pred.hi} times $u$ times that ratio. With every $T$ native the enclosure is at most \V{en.h.tn.max}; with every $X_{21}$ native there is no relative bound.

\begin{table*}[t]
\caption{Grades of the TRTRI products, the inverse's error and the certified enclosure of $\|\widehat Y -
(LL^T)^{-1}\|_F/\|(LL^T)^{-1}\|_F$ (H100, $n = \V{gr.n}$). On $X_{21}$, $r_A/k \le \V{gr.fixed.xmax}$ throughout. n.r.b.: no relative bound.}
\label{tab:grade}
\centering\scriptsize
\setlength{\tabcolsep}{2pt}
\begin{tabular*}{\textwidth}{@{\extracolsep{\fill}}lcccccccccc@{}}
\toprule
 & & \multicolumn{3}{c}{$g = 0$} & \multicolumn{3}{c}{$g = 32$} & \multicolumn{3}{c}{$g = 64$} \\
route & $r_B/k$, all $g$ & $r_A/k$ on $T$ & $\epsilon_X$ & enclosure & $r_A/k$ on $T$ & $\epsilon_X$ & enclosure & $r_A/k$ on $T$ & $\epsilon_X$ & enclosure \\
\midrule
native & \V{gr.blkcublas.bg} & \V{gr.blkcublas.g0.at} & \V{gr.blkcublas.g0.err} & \V{en.blkcublas.g0.rel} & \V{gr.blkcublas.g32.at} & \V{gr.blkcublas.g32.err} & \V{en.blkcublas.g32.rel} & \V{gr.blkcublas.g64.at} & \V{gr.blkcublas.g64.err} & \V{en.blkcublas.g64.rel} \\
DEFAULT & \V{gr.blkcublasemu.bg} & \V{gr.blkcublasemu.g0.at} & \V{gr.blkcublasemu.g0.err} & \V{en.blkcublasemu.g0.rel} & \V{gr.blkcublasemu.g32.at} & \V{gr.blkcublasemu.g32.err} & \V{en.blkcublasemu.g32.rel} & \V{gr.blkcublasemu.g64.at} & \V{gr.blkcublasemu.g64.err} & \V{en.blkcublasemu.g64.rel} \\
EAGER & \V{gr.blkcublasemueager.bg} & \V{gr.blkcublasemueager.g0.at} & \V{gr.blkcublasemueager.g0.err} & \V{en.blkcublasemueager.g0.rel} & \V{gr.blkcublasemueager.g32.at} & \V{gr.blkcublasemueager.g32.err} & \V{en.blkcublasemueager.g32.rel} & \V{gr.blkcublasemueager.g64.at} & \V{gr.blkcublasemueager.g64.err} & \V{en.blkcublasemueager.g64.rel} \\
\midrule
$q$=14 & \V{gr.blkgemmul8m14.bg} & \V{gr.blkgemmul8m14.g0.at} & \V{gr.blkgemmul8m14.g0.err} & \V{en.blkgemmul8m14.g0.rel} & \V{gr.blkgemmul8m14.g32.at} & \V{gr.blkgemmul8m14.g32.err} & \V{en.blkgemmul8m14.g32.rel} & \V{gr.blkgemmul8m14.g64.at} & \V{gr.blkgemmul8m14.g64.err} & \V{en.blkgemmul8m14.g64.rel} \\
$q$=16 & \V{gr.blkgemmul8m16.bg} & \V{gr.blkgemmul8m16.g0.at} & \V{gr.blkgemmul8m16.g0.err} & \V{en.blkgemmul8m16.g0.rel} & \V{gr.blkgemmul8m16.g32.at} & \V{gr.blkgemmul8m16.g32.err} & \V{en.blkgemmul8m16.g32.rel} & \V{gr.blkgemmul8m16.g64.at} & \V{gr.blkgemmul8m16.g64.err} & \V{en.blkgemmul8m16.g64.rel} \\
$q$=20 & \V{gr.blkgemmul8m20.bg} & \V{gr.blkgemmul8m20.g0.at} & \V{gr.blkgemmul8m20.g0.err} & \V{en.blkgemmul8m20.g0.rel} & \V{gr.blkgemmul8m20.g32.at} & \V{gr.blkgemmul8m20.g32.err} & \V{en.blkgemmul8m20.g32.rel} & \V{gr.blkgemmul8m20.g64.at} & \V{gr.blkgemmul8m20.g64.err} & \V{en.blkgemmul8m20.g64.rel} \\
F55 & \V{gr.blkcublasemufixed55.bg} & \V{gr.blkcublasemufixed55.g0.at} & \V{gr.blkcublasemufixed55.g0.err} & \V{en.blkcublasemufixed55.g0.rel} & \V{gr.blkcublasemufixed55.g32.at} & \V{gr.blkcublasemufixed55.g32.err} & \V{en.blkcublasemufixed55.g32.rel} & \V{gr.blkcublasemufixed55.g64.at} & \V{gr.blkcublasemufixed55.g64.err} & \V{en.blkcublasemufixed55.g64.rel} \\
F79 & \V{gr.blkcublasemufixed79.bg} & \V{gr.blkcublasemufixed79.g0.at} & \V{gr.blkcublasemufixed79.g0.err} & \V{en.blkcublasemufixed79.g0.rel} & \V{gr.blkcublasemufixed79.g32.at} & \V{gr.blkcublasemufixed79.g32.err} & \V{en.blkcublasemufixed79.g32.rel} & \V{gr.blkcublasemufixed79.g64.at} & \V{gr.blkcublasemufixed79.g64.err} & \V{en.blkcublasemufixed79.g64.rel} \\
\midrule
$q$=14, eq. & \V{gr.blkeqgemmul8m14.bg} & \V{gr.blkeqgemmul8m14.g0.at} & \V{gr.blkeqgemmul8m14.g0.err} & \V{en.blkeqgemmul8m14.g0.rel} & \V{gr.blkeqgemmul8m14.g32.at} & \V{gr.blkeqgemmul8m14.g32.err} & \V{en.blkeqgemmul8m14.g32.rel} & \V{gr.blkeqgemmul8m14.g64.at} & \V{gr.blkeqgemmul8m14.g64.err} & \V{en.blkeqgemmul8m14.g64.rel} \\
F55, eq. & \V{gr.blkeqcublasemufixed55.bg} & \V{gr.blkeqcublasemufixed55.g0.at} & \V{gr.blkeqcublasemufixed55.g0.err} & \V{en.blkeqcublasemufixed55.g0.rel} & \V{gr.blkeqcublasemufixed55.g32.at} & \V{gr.blkeqcublasemufixed55.g32.err} & \V{en.blkeqcublasemufixed55.g32.rel} & \V{gr.blkeqcublasemufixed55.g64.at} & \V{gr.blkeqcublasemufixed55.g64.err} & \V{en.blkeqcublasemufixed55.g64.rel} \\
\midrule
$q$=14, $T$ native & & & & & & & & & \V{en.blkgemmul8m14Tn.g64.err} & \V{en.blkgemmul8m14Tn.g64.rel} \\
$q$=14, $X_{21}$ native & & & & & & & & & \V{en.blkgemmul8m14Xn.g64.err} & \V{en.blkgemmul8m14Xn.g64.rel} \\
F55, $T$ native & & & & & & & & & \V{en.blkcublasemufixed55Tn.g64.err} & \V{en.blkcublasemufixed55Tn.g64.rel} \\
F55, $X_{21}$ native & & & & & & & & & \V{en.blkcublasemufixed55Xn.g64.err} & \V{en.blkcublasemufixed55Xn.g64.rel} \\
\bottomrule
\end{tabular*}
\end{table*}

\paragraph{Four GPUs}
Table~\ref{tab:gpus} gives every route on the four GPUs and at $n = 16384$. The bits lost per unit of $g$ range from
\V{h100.fx.m14.slope} for 14 moduli to \V{h100.fx.f79.slope} for FIXED 79, and the onset is consistent with the
single-product analysis~\cite{fasi2026}; more bits lower the error, FIXED 111 to
within a small factor of native at $g = 64$. The slopes on the H100 differ from the RTX 4090's by at most
\V{both.fx.slope.maxdiff3} bits per unit of $g$, and over seeds 0 to 4 at $n = 4096$ by at most
\V{seed.slope.maxspread} (H100) and \V{seed4090.slope.maxspread} (RTX 4090). Seeds 1 to 4 at $n = 8192$, $g = 64$ give
\V{h100k.s8.blkgemmul8m14.g64.lo} to \V{h100k.s8.blkgemmul8m14.g64.hi} for 14 moduli and
\V{h100k.s8.blkeqcublasemufixed55.g64.lo} to \V{h100k.s8.blkeqcublasemufixed55.g64.hi} for equilibrated FIXED 55 on
the H100; the other GPUs are alike (supplement). In contiguous chunks of exponents 16 moduli give
\V{ord.g64.blocks.8192.gemmul8m16}, and EAGER stays at native accuracy sorted and in chunks
(\V{ord.g64.sorted.8192.cublasemueager}, \V{ord.g64.blocks.8192.cublasemueager}).

Two further checks agree. A normwise audit of grade-C type passes on every TRTRI product at $n = 4096$, $g = 64$
($\|C - C_h\|_F \le \beta u\|A\|_F\|B\|_F$ with $\beta \le \V{bd.all.units}$ against the host's DGEMM, whose bound is
about $\beta = k$). The scaled residual, which needs no reference, separates the routes exactly as $\epsilon_X$ does: at
most \V{d3.h100.g64.res.acc.max} for every accurate full route and at least \V{d3.h100.g64.res.bad.min} for every failing
one on the H100 (\V{d3.r4090.g64.res.acc.max} and \V{d3.r4090.g64.res.bad.min} on the RTX 4090). Bit-for-bit scale
equivariance is a reference-free symptom: native routes passed all \V{r4090.inv.native.count} checks on each GPU; every
raw fixed route with $g > 0$ failed on the RTX 4090, and all but \V{h100.fx.raw.invariant} (route, input) pair on the
H100. EAGER routes failed too (all \V{r4090.inv.eager.count} cases with $g > 0$) while accurate, so a failure flags a
changed path, not an error.

\begin{table*}[t]
\caption{$\epsilon_X$ at $g = 64$, $n = 8192$, on four GPUs, and at $n = 16384$ (H100). Slope: H100; units: mixed units
(RTX 4090); ADP: worst guarded / native ratio over each GPU's grid.}
\label{tab:gpus}
\centering\scriptsize
\setlength{\tabcolsep}{3pt}
\begin{tabular*}{\textwidth}{@{\extracolsep{\fill}}lccccccc@{}}
\toprule
route & slope & H100 & A100 & RTX 4090 & L40S & $n$=16384 & units \\
\midrule
native & \V{h100.fx.native.slope} & \V{h100.fx.native.g64} & \V{a100.fx.native.g64} & \V{r4090.fx.native.g64} & \V{l40s.fx.native.g64} & \V{h100k.n16k.blkcublas.g64} & \V{r4090.fx.native.um} \\
$q$=14 & \V{h100.fx.m14.slope} & \V{h100.fx.m14.g64} & \V{a100.fx.m14.g64} & \V{r4090.fx.m14.g64} & \V{l40s.fx.m14.g64} & \V{h100k.n16k.blkgemmul8m14.g64} & \V{r4090.fx.m14.um} \\
$q$=16 & \V{h100.fx.m16.slope} & \V{h100.fx.m16.g64} & \V{a100.fx.m16.g64} & \V{r4090.fx.m16.g64} & \V{l40s.fx.m16.g64} & \V{h100k.n16k.blkgemmul8m16.g64} & \V{r4090.fx.m16.um} \\
$q$=20 & \V{h100.fx.m20.slope} & \V{h100.fx.m20.g64} & \V{a100.fx.m20.g64} & \V{r4090.fx.m20.g64} & \V{l40s.fx.m20.g64} & \V{h100k.n16k.blkgemmul8m20.g64} & \V{r4090.fx.m20.um} \\
F55 & \V{h100.fx.f55.slope} & \V{h100.fx.f55.g64} & \V{a100.fx.f55.g64} & \V{r4090.fx.f55.g64} & \V{l40s.fx.f55.g64} & \V{h100k.n16k.blkcublasemufixed55.g64} & \V{r4090.fx.f55.um} \\
F79 & \V{h100.fx.f79.slope} & \V{h100.fx.f79.g64} & \V{a100.fx.f79.g64} & \V{r4090.fx.f79.g64} & \V{l40s.fx.f79.g64} & \V{h100k.n16k.blkcublasemufixed79.g64} & \V{r4090.fx.f79.um} \\
F111 & \V{h100.fx.f111.slope} & \V{h100.fx.f111.g64} & \V{a100.fx.f111.g64} & \V{r4090.fx.f111.g64} & \V{l40s.fx.f111.g64} & -- & \V{r4090.fx.f111.um} \\
\midrule
$q$=14, eq. & \V{h100.fx.eqm14.slope} & \V{h100.fx.eqm14.g64} & \V{a100.fx.eqm14.g64} & \V{r4090.fx.eqm14.g64} & \V{l40s.fx.eqm14.g64} & \V{h100k.n16k.blkeqgemmul8m14.g64} & \V{r4090.fx.eqm14.um} \\
$q$=16, eq. & \V{h100.fx.eqm16.slope} & \V{h100.fx.eqm16.g64} & \V{a100.fx.eqm16.g64} & \V{r4090.fx.eqm16.g64} & \V{l40s.fx.eqm16.g64} & -- & \V{r4090.fx.eqm16.um} \\
F55, eq. & \V{h100.fx.eqf55.slope} & \V{h100.fx.eqf55.g64} & \V{a100.fx.eqf55.g64} & \V{r4090.fx.eqf55.g64} & \V{l40s.fx.eqf55.g64} & \V{h100k.n16k.blkeqcublasemufixed55.g64} & \V{r4090.fx.eqf55.um} \\
F79, eq. & \V{h100.fx.eqf79.slope} & \V{h100.fx.eqf79.g64} & \V{a100.fx.eqf79.g64} & \V{r4090.fx.eqf79.g64} & \V{l40s.fx.eqf79.g64} & -- & \V{r4090.fx.eqf79.um} \\
\midrule
ADP & & \V{h100.err.worst.ratio}$\times$ & \V{a100.grid.worst.ratio}$\times$ & \V{r4090.err.worst.ratio}$\times$ & \V{l40s.grid.worst.ratio}$\times$ & & \\
\bottomrule
\end{tabular*}
\end{table*}

Where the difficulty is conditioning rather than scaling, emulation tracks native. With $\kappa = 10^8$, potrf
followed by potri differs from LAPACK by \V{r4090.ctl.cond_1e8.chol} natively and \V{r4090.ctl.cond_1e8.chol.eager}
under EAGER on the RTX 4090 (\V{h100.ctl.cond_1e8.chol} and \V{h100.ctl.cond_1e8.chol.eager} on the H100), a
factor-level difference of order $\kappa u$; the Gaussian-process kernel gives \V{r4090.ctl.gp_opt.chol} both ways. On
the H100's second version, cuBLAS \V{h100.v132.cublas} and cuSOLVER \V{h100.v132.cusolver}, every cuSOLVER error is
within \V{h100.v132.err.reldiff.pct}\% of the pinned version's.

\paragraph{Replacement and re-association}
Table~\ref{tab:replace} extends Fig.~\ref{fig:main}(c) with $n = 8192$, LAUUM and the
16 products of largest and smallest excess. Leaving any single $T$ emulated, with all others native, still gives
\V{h100k.loo.blkgemmul8m14.n8192.lo} or worse at $n = 8192$. On a transparent CPU emulator ($n = \V{ac.n}$, 8 slices) the solve-based inverse of
$DMD$ is bit-for-bit $D^{-1}$ times its inverse of $M$ times $D^{-1}$ at every tested $g$ (\V{ac.s8.inv} of
\V{ac.s8.cases}), with errors from \V{ac.s8.solve.lo} to \V{ac.s8.solve.hi} where the product form reaches
\V{ac.s8.prod.g64}. Through the harness at $n \in \{4096, 8192\}$ and $g$ from 16 to 64 it is bit-for-bit scale
equivariant in \V{as.h100.blksgemmul8m14.inv} of \V{as.h100.blksgemmul8m14.invn} cases with 14 moduli and
\V{as.h100.blkscublasemufixed55.inv} of \V{as.h100.blkscublasemufixed55.invn} with FIXED 55 on the H100, likewise on
the RTX 4090 (\V{as.r4090.blksgemmul8m14.inv} and \V{as.r4090.blkscublasemufixed55.inv} of
\V{as.r4090.blkscublasemufixed55.invn}), and so is every complete device pipeline from cuSOLVER's potrf
(\V{eqv.pipe.solve} of \V{eqv.pipe.solven}; product form: \V{eqv.pipe.prod} of 2). On the H100 its error on the raw
input never exceeds \V{as.h100.blksgemmul8m14.errmax} and \V{as.h100.blkscublasemufixed55.errmax}, where the product
form reaches \V{as.h100.blkgemmul8m14.errmax}; on mixed units it stays at most \V{as.r4090.um.blksgemmul8m14}.

\begin{table}[t]
\caption{$\epsilon_X$ with chosen products native (H100, $g = 64$). Top/bottom 16: largest/smallest scale excess.}
\label{tab:replace}
\centering\scriptsize
\setlength{\tabcolsep}{2pt}
\begin{tabular}{@{}lcccc@{}}
\toprule
 & \multicolumn{2}{c}{GEMMul8 $q = 14$} & \multicolumn{2}{c}{FIXED 55} \\
computed natively & $n$=4096 & $n$=8192 & $n$=4096 & $n$=8192 \\
\midrule
none & \V{rp.blkgemmul8m14.4096.none} & \V{rp.blkgemmul8m14.8192.none} & \V{rp.blkcublasemufixed55.4096.none} & \V{rp.blkcublasemufixed55.8192.none} \\
LAUUM & \V{rp.blkgemmul8m14.4096.lauum_only} & \V{rp.blkgemmul8m14.8192.lauum_only} & \V{rp.blkcublasemufixed55.4096.lauum_only} & \V{rp.blkcublasemufixed55.8192.lauum_only} \\
every $X_{21}$ & \V{rp2.blkgemmul8m14.4096.outer_all} & \V{rp2.blkgemmul8m14.8192.outer_all} & \V{rp2.blkcublasemufixed55.4096.outer_all} & \V{rp2.blkcublasemufixed55.8192.outer_all} \\
bottom 16 & \V{rp.blkgemmul8m14.4096.bottom16} & \V{rp.blkgemmul8m14.8192.bottom16} & \V{rp.blkcublasemufixed55.4096.bottom16} & \V{rp.blkcublasemufixed55.8192.bottom16} \\
top 16 & \V{rp.blkgemmul8m14.4096.top16} & \V{rp.blkgemmul8m14.8192.top16} & \V{rp.blkcublasemufixed55.4096.top16} & \V{rp.blkcublasemufixed55.8192.top16} \\
every $T$ but one & $\ge$\V{h100k.loo.blkgemmul8m14.n4096.lo} & $\ge$\V{h100k.loo.blkgemmul8m14.n8192.lo} & $\ge$\V{h100k.loo.blkcublasemufixed55.n4096.lo} & $\ge$\V{h100k.loo.blkcublasemufixed55.n8192.lo} \\
every $T$ & \V{rp2.blkgemmul8m14.4096.inner_all} & \V{rp2.blkgemmul8m14.8192.inner_all} & \V{rp2.blkcublasemufixed55.4096.inner_all} & \V{rp2.blkcublasemufixed55.8192.inner_all} \\
\bottomrule
\end{tabular}
\end{table}

\paragraph{Real matrices and library versions}
The 16 SuiteSparse matrices are those with $1024 \le n \le 8192$ and diagonal span at least $2^{20}$ that the NIST %
Matrix Market mirrors, chosen by a rule fixed before any matrix was seen; the covariances use the 1990 California %
census (seven variables) and the UCI air-quality data (twelve), in raw units. Table~\ref{tab:real} gives the group
maxima. Equilibrated FIXED 55 stays within a factor \V{l40s.realsum.eqf55.max} of native on every matrix. Equilibrated 16 moduli stay within a factor \V{l40s.realsum.eqm16.max} of native on every matrix;
equilibrated 14 moduli stay \V{a100.realsum.eqm14.max} times native on bcsstk23 on the A100 and at most
\V{l40s.realsum.eqm14.second} times on every other matrix (\V{a100.realsum.eqm14.second} on the A100). Over the
\V{l40s.vs.nver} releases the guarded routes' largest factor over native is \V{l40s.vs.guard.max} on the L40S,
\V{a100.vs.guard.max} on the A100 and \V{h100k.vs.guard.max} on the H100. Equilibrated FIXED 55 reaches native level from cuBLAS
13.7 (\V{l40s.vs.v137027.blkeqcublasemufixed55.g64} on the L40S) and stops at %
\V{l40s.vs.v13602.blkeqcublasemufixed55.g64} before it; on the H100 the change arrives in 13.8 %
(\V{h100k.vs.v137027.blkeqcublasemufixed55.g64} under 13.7, \V{h100k.vs.v13804.blkeqcublasemufixed55.g64} under 13.8). %
cuSOLVER's \texttt{potri} under its own FIXED control gives \V{l40s.csfx.b55.potri.g64}, \V{l40s.csfx.b79.potri.g64}
and \V{l40s.csfx.b111.potri.g64} with 55, 79 and 111 bits on the L40S and \V{x4090.csfx.b55.potri.g64} with 55 on the
RTX 4090; equilibrated, \V{l40s.csfx.b55.potrieq.g64}. The supplement gives every matrix and every release.

\begin{table*}[t]
\caption{Real SPD matrices: largest error per group in the equilibrated frame (L40S; last column A100). Span:
$\log_2$ of the diagonal span. ADP: cuSOLVER EAGER.}
\label{tab:real}
\centering\scriptsize
\setlength{\tabcolsep}{1.5pt}
\begin{tabular*}{\textwidth}{@{\extracolsep{\fill}}lccccccccc@{}}
\toprule
 & & & & & & \multicolumn{4}{c}{equilibrated} \\
matrices & span & native & ADP & $q$=14 & F55 & $q$=14 & $q$=16 & F55 & $q$=14, A100 \\
\midrule
census (4) & \V{rm.cal7vmm585.span} & \V{l40s.realsum.census.native.max} & \V{l40s.realsum.census.eager.max} & \V{l40s.realsum.census.m14.max} & \V{l40s.realsum.census.f55.max} & \V{l40s.realsum.census.eqm14.max} & \V{l40s.realsum.census.eqm16.max} & \V{l40s.realsum.census.eqf55.max} & \V{a100.realsum.census.eqm14.max} \\
air (4) & \V{rm.air12vmm341.span} & \V{l40s.realsum.air.native.max} & \V{l40s.realsum.air.eager.max} & \V{l40s.realsum.air.m14.max} & \V{l40s.realsum.air.f55.max} & \V{l40s.realsum.air.eqm14.max} & \V{l40s.realsum.air.eqm16.max} & \V{l40s.realsum.air.eqf55.max} & \V{a100.realsum.air.eqm14.max} \\
SuiteSparse (\V{rm.ss.count}) & \V{rm.ss.span.lo}--\V{rm.ss.span.hi} & \V{l40s.realsum.ss.native.max} & \V{l40s.realsum.ss.eager.max} & \V{l40s.realsum.ss.m14.max} & \V{l40s.realsum.ss.f55.max} & \V{l40s.realsum.ss.eqm14.max} & \V{l40s.realsum.ss.eqm16.max} & \V{l40s.realsum.ss.eqf55.max} & \V{a100.realsum.ss.eqm14.max} \\
\enspace bcsstk23 & \V{rm.bcsstk23.span} & \V{l40s.real.ssbcsstk23.blkcublas.sc} & \V{l40s.real.ssbcsstk23.cusolveremueagerpotri.sc} & \V{l40s.real.ssbcsstk23.blkgemmul8m14.sc} & \V{l40s.real.ssbcsstk23.blkcublasemufixed55.sc} & \V{l40s.real.ssbcsstk23.blkeqgemmul8m14.sc} & \V{l40s.real.ssbcsstk23.blkeqgemmul8m16.sc} & \V{l40s.real.ssbcsstk23.blkeqcublasemufixed55.sc} & \V{a100.real.ssbcsstk23.blkeqgemmul8m14.sc} \\
\bottomrule
\end{tabular*}
\end{table*}

\section{The guard: estimates, switch and DEFAULT}
\label{app:switch}
\label{sec:switch}
\label{app:ptr}

\paragraph{An explanatory proxy}
For $C = AB$ with $c_{ij} \ne 0$ and $r_i$, $s_j$, $p_{ij}$ as in Section~\ref{sec:background}, the excess $e_{ij} =
\log_2(r_is_j/|c_{ij}|)$ is the sum of $\log_2(r_is_j/p_{ij})$, the ratio of the grade-B to the grade-A right-hand side,
and $\log_2(p_{ij}/|c_{ij}|)$, ordinary cancellation. The harness records its 99th percentile $e_{99}$ and maximum per
product. It needs the product, so it explains a library's choice after the fact and is not an online estimate.
The fit of Fig.~\ref{fig:bits} covers all \V{r4090.adp.calls} TRTRI products of the scaled grid on both GPUs; at $n = 8192$
they report \V{r4090.adp.bits.g0.lo} to \V{r4090.adp.bits.g0.hi} bits at $g = 0$. The figure plots every TRTRI product under EAGER at $n \in \{2048, 4096, 8192\}$, $g \in \{0, 16, 32, 48,
64\}$, raw and equilibrated, on the RTX 4090; the H100's estimates are identical. Componentwise, no ADP route's error exceeds its native counterpart's by more than a factor of
\V{r4090.cw.worst.ratio} (RTX 4090) or \V{h100.cw.worst.ratio} (H100) over the grid.

\paragraph{Where a product switches path}
cuBLAS writes its bit estimate to an output pointer the caller provides. The pointer holds the estimate whether or not
the product emulates (cuBLASLt documents $-1$ when emulation is not used), so dispatch has to be read from outputs, time
and kernels. The probes of Fig.~\ref{fig:dispatch} compare each output bit for bit with native DGEMM. The estimate is \V{r4090.probe.s0.bits} bits at zero span and
\V{r4090.probe.s200.bits} at a span of 200 bits on both GPUs.
On the RTX 4090 with cuBLAS \V{r4090.cublas}, outputs of $1024 \times 1024$ or smaller step between \V{r4090.step.small.below} and \V{r4090.step.small.above} bits (for example from
\V{r4090.step.small.t48} to \V{r4090.step.small.t52} times native), $2048 \times 2048$ outputs between
\V{r4090.thr.2048.emu} and \V{r4090.thr.2048.cap} bits (from at most \V{r4090.thr.2048.emutime.hi} to at least
\V{r4090.thr.2048.captime.lo} times native). Above the step, outputs of $2048 \times 2048$ and larger are
bit-identical to native DGEMM on the RTX 4090; smaller ones are not, so the path above the step is not always the default DGEMM kernel.
On the H100, Nsight Systems traces show the same kernels launched on both sides of the switch, with the durations
moving: for the $4096^3$ product the INT8 GEMMs take \V{ko.4096.s44.int8us}~$\mu$s per call at \V{kern.4096.s44.bits}
bits and \V{ko.4096.s48.int8us}~$\mu$s at \V{kern.4096.s48.bits}. Above the switch a CUTLASS FP64 tensor-core GEMM of
the emulation library does the product, not cuBLAS's default FP64 kernel, which accounts for over-cap outputs that are
accurate but not bit-identical to native DGEMM. The $4096^3$ product takes \V{kern.4096.s44.ratio} times native at
\V{kern.4096.s44.bits} bits and \V{kern.4096.s48.ratio} at \V{kern.4096.s48.bits}, which places the H100's switch where
the RTX 4090's is. Exact zeros on both operands, as in a single-GEMM LAUUM, also change the path: the $8192^3$ product
returns native's output at \V{r4090.probe.tri_8192.time} times native time on the RTX 4090.

\paragraph{Inside cuSOLVER and under DEFAULT}
cuSOLVER 12.3.5 imports eleven FP64 level-3 entry points from cuBLAS, and an \texttt{LD\_PRELOAD} interposer wraps all
of them, logging each internal call's operation, shape, strategy and estimate. With all eleven wrapped, potri at
$n = 4096$ and $g = 64$ makes \V{spyc.eager.calls} internal FP64 level-3 calls in every mode; cuBLAS's own API log of
the same call, taken without the interposer, records \V{apilog.native.l3}, \V{apilog.default.l3} and
\V{apilog.eager.l3} under native, DEFAULT and EAGER, the interposer's count operation for operation. At $n = 8192$,
$g = 64$, the internal TRSM calls report \V{r4090.spy.eager.trsm.lo} to \V{r4090.spy.eager.trsm.hi} bits under EAGER
and TRMM up to \V{r4090.spy.eager.trmm.hi}, identically on both GPUs; none of the \V{r4090.spy.eager.syrk.calls} SYRK
calls is emulated, as documented below compute capability 10.0. DEFAULT, the strategy a caller gets without asking
for EAGER, decides for itself. On the RTX 4090 it writes an estimate on \V{dc4090.written} of the
\V{r4090.default.calls.all} GEMMs issued by the blocked routes: \V{dc4090.below} at or below \V{dc4090.lo} bits,
emulated at every tested output size, \V{dc4090.above} at or above \V{dc4090.hi} bits, native at every size, and
\V{dc4090.mid} in between, unclassified; the path is inferred from the estimate against the switch. \ifnum\V{dt4090.calls}>0{}One
complete RTX 4090 DEFAULT execution, traced kernel by kernel with CUPTI, took the inferred path on \V{dt4090.time} of
its \V{dt4090.calls} calls. \fi{}On the H100 DEFAULT writes no estimate on any blocked GEMM
(\V{h100.default.emulated.all} of \V{h100.default.calls.all}) or internal cuSOLVER call, and kernel traces of potri show
emulation kernels on \V{pk.default.emu} of \V{pk.default.all} calls. At $n = 8192$, $g = 64$, DEFAULT emulates none of
\V{h100.spy.calls} internal potri calls on the H100 and A100 but \V{l40s.spy.default.written} on the L40S and RTX 4090, and
EAGER \V{l40s.spy.eager.written} on all four. A caller who enables emulation through cuSOLVER's strategy setting thus gets
a different numerical path on each GPU.

\paragraph{The bit-count pointer on cuBLAS 13.3}
On the H100's second version, directing the bit-count output to page-locked host memory, as the harness does, slows
single EAGER products to \V{v132.thr.time.lo} to \V{v132.thr.time.hi} times native time. A controlled experiment places
the added time in the emulation library's epilogue kernel, launched with the same configuration and running
\V{ed.ratio} times longer with the pointer; cuSOLVER does not enable the pointer on its internal calls. Instrumenting an
opaque library can thus perturb what is measured (supplement).

\section{Cost}
\label{app:cost}

The times below describe the recorded sessions; they do not bound what another device, driver, clock state or session
would give, and the conclusions rest on the order of the routes and on ratios far from 1. Table~\ref{tab:full} lists every route. All \V{pr.h100.nresolved} matched comparisons on the H100
(\V{pr.h100.sessions} sessions on \V{pr.h100.nodes} nodes) and all \V{pr.r4090.nresolved} on the RTX 4090 are resolved
by 97.9\% order-statistic intervals under a rule fixed before computing; the closest to 1 is equilibration's cost on %
the H100, session medians \V{pr.h100.eqcost.mlo} to \V{pr.h100.eqcost.mhi}. Blocking LAUUM saves time on every GPU
(blocked / one GEMM, native: \V{h100k.d3x.lauum} on the H100, \V{x4090.d3x.lauum} on the RTX 4090), and equilibrated
FIXED 55 takes \V{h100k.d3x.eq2f55} of the equilibrated native route's time on the H100 with blocked LAUUM and
\V{h100k.d3x.eqf55} with one-GEMM LAUUM (\V{x4090.d3x.eq2f55} and \V{x4090.d3x.eqf55} on the RTX 4090).

On the RTX 4090 the fastest accurate route, equilibrated GEMMul8 with 16 moduli and blocked LAUUM, inverts an $8192
\times 8192$ matrix in \V{d3.r4090.sess.blk2eqgemmul8m16.ms.lo} to \V{d3.r4090.sess.blk2eqgemmul8m16.ms.hi}~ms over
sessions, against \V{d3.r4090.sess.cusolver.ms.lo} to \V{d3.r4090.sess.cusolver.ms.hi}~ms for cuSOLVER. On the H100
the fastest accurate route is native, the blocked inversion with blocked LAUUM, and the fastest accurate emulated
route, equilibrated FIXED 55 with blocked LAUUM, takes \V{fair.h100.emutwo} of its time. With upload and download from
pageable memory inside the clock the margins shrink: the RTX 4090's fastest accurate route takes
\V{d3.r4090.g64.blk2eqgemmul8m16.xratio} of cuSOLVER's time, and on the H100 the copies dominate (native cuSOLVER
\V{d3.h100.sess.cusolver.xms.lo} to \V{d3.h100.sess.cusolver.xms.hi}~ms). Emulation also needs workspace: the sampled
peak of device memory reaches \V{d3.h100.g64.blkgemmul8m14.mem}~MiB for 14 moduli and
\V{d3.h100.g64.blkcublasemufixed55.mem}~MiB for FIXED 55 on the H100, against \V{d3.h100.g64.blkcublas.mem}~MiB native.
The $T$ products are half of TRTRI's GEMM work, so the $T$-native hybrid gives back part of emulation's gain where FP64
is slow: \V{fair.r4090.tn} times the equilibrated route's time on the RTX 4090, and the solve-based route
\V{fair.r4090.solve}. Inside cuSOLVER's potri on an existing factor, the spread raises DEFAULT on the RTX 4090 from
\V{tm4090.8192.scaleg0.emu} of native time at $g = 0$ to \V{tm4090.8192.scaleg64.emu} at $g = 64$, and equilibration
recovers it (EAGER \V{tm4090.8192.scaleg64.emueagereq} on the equilibrated input).

\begin{table*}[t]
\caption{Full routes, device-resident, $n = 8192$, $g = 64$: median time in ms [interquartile range], ratio to
native cuSOLVER and $\epsilon_X$. All columns, rows and sessions: supplement.}
\label{tab:full}
\centering\scriptsize
\setlength{\tabcolsep}{2pt}
\begin{tabular*}{\textwidth}{@{\extracolsep{\fill}}lcccccc@{}}
\toprule
 & \multicolumn{3}{c}{RTX 4090} & \multicolumn{3}{c}{H100} \\
\cmidrule(lr){2-4}\cmidrule(lr){5-7} %
route & ms [IQR] & ratio & $\epsilon_X$ & ms [IQR] & ratio & $\epsilon_X$ \\
\midrule
cuSOLVER, native & \V{d3.r4090.g64.cusolver.ms} [\V{d3.r4090.g64.cusolver.iqr}] & \V{d3.r4090.g64.cusolver.ratio} & \V{d3.r4090.g64.cusolver.err} & \V{d3.h100.g64.cusolver.ms} [\V{d3.h100.g64.cusolver.iqr}] & \V{d3.h100.g64.cusolver.ratio} & \V{d3.h100.g64.cusolver.err} \\
cuSOLVER, DEFAULT & \V{d3.r4090.g64.cusolveremu.ms} [\V{d3.r4090.g64.cusolveremu.iqr}] & \V{d3.r4090.g64.cusolveremu.ratio} & \V{d3.r4090.g64.cusolveremu.err} & \V{d3.h100.g64.cusolveremu.ms} [\V{d3.h100.g64.cusolveremu.iqr}] & \V{d3.h100.g64.cusolveremu.ratio} & \V{d3.h100.g64.cusolveremu.err} \\
cuSOLVER, EAGER & \V{d3.r4090.g64.cusolveremueager.ms} [\V{d3.r4090.g64.cusolveremueager.iqr}] & \V{d3.r4090.g64.cusolveremueager.ratio} & \V{d3.r4090.g64.cusolveremueager.err} & \V{d3.h100.g64.cusolveremueager.ms} [\V{d3.h100.g64.cusolveremueager.iqr}] & \V{d3.h100.g64.cusolveremueager.ratio} & \V{d3.h100.g64.cusolveremueager.err} \\
\midrule
blocked, native, blocked LAUUM & \V{d3.r4090.g64.blk2cublas.ms} [\V{d3.r4090.g64.blk2cublas.iqr}] & \V{d3.r4090.g64.blk2cublas.ratio} & \V{d3.r4090.g64.blk2cublas.err} & \V{d3.h100.g64.blk2cublas.ms} [\V{d3.h100.g64.blk2cublas.iqr}] & \V{d3.h100.g64.blk2cublas.ratio} & \V{d3.h100.g64.blk2cublas.err} \\
\midrule
$q$=14 & \V{d3.r4090.g64.blkgemmul8m14.ms} [\V{d3.r4090.g64.blkgemmul8m14.iqr}] & \V{d3.r4090.g64.blkgemmul8m14.ratio} & \V{d3.r4090.g64.blkgemmul8m14.err} & \V{d3.h100.g64.blkgemmul8m14.ms} [\V{d3.h100.g64.blkgemmul8m14.iqr}] & \V{d3.h100.g64.blkgemmul8m14.ratio} & \V{d3.h100.g64.blkgemmul8m14.err} \\
$q$=14, $T$ native & \V{d3.r4090.g64.blkgemmul8m14Tn.ms} [\V{d3.r4090.g64.blkgemmul8m14Tn.iqr}] & \V{d3.r4090.g64.blkgemmul8m14Tn.ratio} & \V{d3.r4090.g64.blkgemmul8m14Tn.err} & \V{d3.h100.g64.blkgemmul8m14Tn.ms} [\V{d3.h100.g64.blkgemmul8m14Tn.iqr}] & \V{d3.h100.g64.blkgemmul8m14Tn.ratio} & \V{d3.h100.g64.blkgemmul8m14Tn.err} \\
$q$=14, eq. & \V{d3.r4090.g64.blkeqgemmul8m14.ms} [\V{d3.r4090.g64.blkeqgemmul8m14.iqr}] & \V{d3.r4090.g64.blkeqgemmul8m14.ratio} & \V{d3.r4090.g64.blkeqgemmul8m14.err} & \V{d3.h100.g64.blkeqgemmul8m14.ms} [\V{d3.h100.g64.blkeqgemmul8m14.iqr}] & \V{d3.h100.g64.blkeqgemmul8m14.ratio} & \V{d3.h100.g64.blkeqgemmul8m14.err} \\
$q$=14, solve-based & \V{d3.r4090.g64.blksgemmul8m14.ms} [\V{d3.r4090.g64.blksgemmul8m14.iqr}] & \V{d3.r4090.g64.blksgemmul8m14.ratio} & \V{d3.r4090.g64.blksgemmul8m14.err} & \V{d3.h100.g64.blksgemmul8m14.ms} [\V{d3.h100.g64.blksgemmul8m14.iqr}] & \V{d3.h100.g64.blksgemmul8m14.ratio} & \V{d3.h100.g64.blksgemmul8m14.err} \\
\midrule
$q$=16, eq. & \V{d3.r4090.g64.blkeqgemmul8m16.ms} [\V{d3.r4090.g64.blkeqgemmul8m16.iqr}] & \V{d3.r4090.g64.blkeqgemmul8m16.ratio} & \V{d3.r4090.g64.blkeqgemmul8m16.err} & \V{d3.h100.g64.blkeqgemmul8m16.ms} [\V{d3.h100.g64.blkeqgemmul8m16.iqr}] & \V{d3.h100.g64.blkeqgemmul8m16.ratio} & \V{d3.h100.g64.blkeqgemmul8m16.err} \\
$q$=16, eq., blocked LAUUM & \V{d3.r4090.g64.blk2eqgemmul8m16.ms} [\V{d3.r4090.g64.blk2eqgemmul8m16.iqr}] & \V{d3.r4090.g64.blk2eqgemmul8m16.ratio} & \V{d3.r4090.g64.blk2eqgemmul8m16.err} & \V{d3.h100.g64.blk2eqgemmul8m16.ms} [\V{d3.h100.g64.blk2eqgemmul8m16.iqr}] & \V{d3.h100.g64.blk2eqgemmul8m16.ratio} & \V{d3.h100.g64.blk2eqgemmul8m16.err} \\
\midrule
F55 & \V{d3.r4090.g64.blkcublasemufixed55.ms} [\V{d3.r4090.g64.blkcublasemufixed55.iqr}] & \V{d3.r4090.g64.blkcublasemufixed55.ratio} & \V{d3.r4090.g64.blkcublasemufixed55.err} & \V{d3.h100.g64.blkcublasemufixed55.ms} [\V{d3.h100.g64.blkcublasemufixed55.iqr}] & \V{d3.h100.g64.blkcublasemufixed55.ratio} & \V{d3.h100.g64.blkcublasemufixed55.err} \\
F55, $T$ native & \V{d3.r4090.g64.blkcublasemufixed55Tn.ms} [\V{d3.r4090.g64.blkcublasemufixed55Tn.iqr}] & \V{d3.r4090.g64.blkcublasemufixed55Tn.ratio} & \V{d3.r4090.g64.blkcublasemufixed55Tn.err} & \V{d3.h100.g64.blkcublasemufixed55Tn.ms} [\V{d3.h100.g64.blkcublasemufixed55Tn.iqr}] & \V{d3.h100.g64.blkcublasemufixed55Tn.ratio} & \V{d3.h100.g64.blkcublasemufixed55Tn.err} \\
F55, eq., blocked LAUUM & \V{d3.r4090.g64.blk2eqcublasemufixed55.ms} [\V{d3.r4090.g64.blk2eqcublasemufixed55.iqr}] & \V{d3.r4090.g64.blk2eqcublasemufixed55.ratio} & \V{d3.r4090.g64.blk2eqcublasemufixed55.err} & \V{d3.h100.g64.blk2eqcublasemufixed55.ms} [\V{d3.h100.g64.blk2eqcublasemufixed55.iqr}] & \V{d3.h100.g64.blk2eqcublasemufixed55.ratio} & \V{d3.h100.g64.blk2eqcublasemufixed55.err} \\
F55, solve-based & \V{d3.r4090.g64.blkscublasemufixed55.ms} [\V{d3.r4090.g64.blkscublasemufixed55.iqr}] & \V{d3.r4090.g64.blkscublasemufixed55.ratio} & \V{d3.r4090.g64.blkscublasemufixed55.err} & \V{d3.h100.g64.blkscublasemufixed55.ms} [\V{d3.h100.g64.blkscublasemufixed55.iqr}] & \V{d3.h100.g64.blkscublasemufixed55.ratio} & \V{d3.h100.g64.blkscublasemufixed55.err} \\
\bottomrule
\end{tabular*}
\end{table*}

\section{Application}
\label{app:cert}
\label{app:proj}

\paragraph{Design}
With $B_{\mathrm{cov}}$ the covariance of the five variables (median age, total rooms, population, median income and
median house value) and $C$ a squared-exponential kernel on location, the model's covariance is $B_{\mathrm{cov}}
\otimes (C + \tau I)$. If house value is observed at every training site, the model's secondary weights are exactly
zero (runs of that kind are in the artifact as a negative control). The design is therefore heterotopic: the four
secondary variables are observed at all $m$ training sites and house value at a predetermined half of them, so the
covariance has order $n = 4.5m$, $n \in \{4608, 9216\}$, in its raw units. $B_{\mathrm{cov}}$ and the means come from
every block group except the 256 held-out ones, so the covariance model is fixed across designs. The length scales
are 0.1 and 0.3 degrees (about 11 and 33 km); \V{ckh.dups} training sites share coordinates with another, and the %
nugget keeps $K$ positive definite. At the model's nugget the secondary observations contribute \V{ckh.t1.sec.min} to
\V{ckh.t1.sec.max} dollars RMS to the prediction, and cokriging changes the held-out RMSE by \V{ckh.t1.gain.min} to
\V{ckh.t1.gain.max}\%. Power-of-two equilibration reduces $\kappa_2(K)$ from \V{ckh.t1.cond.min}--\V{ckh.t1.cond.max}
to at most \V{ckh.t1.condeq.max}, and the TRTRI products' 99th-percentile scale excess reaches
\V{ckh.lm.l3.t1.4608.escale} bits ($n = 4608$). At $n = 4608$ (Table~\ref{tab:appm}) the reference predictor's held-out RMSE is \V{ckh.lm.l3.t1.4608.rmse},
\V{ckh.lm.l3.t3.4608.rmse} and \V{ckh.lm.l3.t6.4608.rmse} dollars at the three nuggets.
At $\tau = 10^{-6}$ the binary64 eigensolver returns a negative smallest eigenvalue estimate for \V{ckh.t6.indefinite}
of the \V{ckh.t6.count} matrices, which are positive definite in exact arithmetic; that nugget is a numerical stress
case only, and nothing is certified there.

\paragraph{Results}
Every ADP route stays at native's level against the reference (at most \V{ckh.eager.worst} of the held-out RMSE for
EAGER, \V{ckh.native.worst} for native). Over both sizes the weakest fixed routes reach \V{ckh.t1.fixed.worst} of the
held-out RMSE at the model's nugget, \V{ckh.t3.fixed.worst} at $\tau = 10^{-3}$ and \V{ckh.t6.fixed.worst} in the
stress case; equilibration brings every fixed route to at most \V{ckh.fixedeq.worst}. Recalibrating $B_{\mathrm{cov}}$
and the means on the observed house values only (\V{ob.cases} cases) changes the held-out RMSE from
\V{ob.t1.rmse.base} to \V{ob.t1.rmse} dollars at $\tau = 10^{-1}$ and leaves these findings in place: the worst raw
fixed route moves from \V{ob.t1.fixed.base} to \V{ob.t1.fixed} and equilibrated routes stay at or below
\V{ob.t3.fixedeq}.

\paragraph{The reference certificate}
The reference prediction is computed by binary64 Cholesky of $K$ with iterative refinement, residuals and iterates in
x86 long double; for all \V{rc.cases} systems ($n = 4608$) the componentwise backward error falls to at most
\V{rc.eta.max} within \V{rc.steps} steps. \emph{Exact} below means exact arithmetic on the stored binary64 arrays $K$,
$b$, $k_j$: $x = K^{-1}b$ and $s_j = k_j^Tx$. Let $\hat s_j$ be the computed prediction, $r = b - K\hat x$ the exact
residual of the computed solution, $S$ the power-of-two equilibration and $0 < \ell \le \lambda_{\min}(SKS)$. Since
$\hat x - x = -S(SKS)^{-1}Sr$,
\begin{equation}
|\hat s_j - s_j| \le \|Sk_j\|_2\|Sr\|_2/\ell + \eta_j,
\label{eq:c1}
\end{equation}
where $\eta_j = \gamma_n|k_j|^T|\hat x| + u|\hat s_j|$ encloses the rounding of the long-double dot product and its
conversion. With $K_* = P[(C_* + \tau I) \otimes B_{\mathrm{cov}}]P^T$ the exact model matrix, $S_vB_{\mathrm{cov}}S_v
\succeq \mu I$ proved by an exact-rational $LDL^T$, and $\Delta \ge \|S(K - K_*)S\|_F$ bounding the stored-versus-model
rounding, Weyl's inequality gives
\begin{equation}
\lambda_{\min}(SKS) \ge \tau\mu - \Delta =: \ell.
\label{eq:c2}
\end{equation}
With $B_{\mathrm{cert}}$ an upper bound on the RMS over the held-out points of the right-hand side of~\eqref{eq:c1}, $R$
a lower bound on the reference's held-out RMSE and $d$ an upper bound on the RMS discrepancy of a route's predictions
from the reference,
\begin{equation}
\frac{\mathrm{RMS}(\hat s_{\mathrm{route}} - s)}{\mathrm{RMSE}_{\mathrm{exact}}} \le \frac{d +
B_{\mathrm{cert}}}{R - B_{\mathrm{cert}}}, \quad R > B_{\mathrm{cert}}.
\label{eq:c3}
\end{equation}
Here $\mu \ge \V{rb.mu}$, $\Delta \le \V{rb.delta}$ and $\ell/\tau \ge \V{rb.lratio}$ in all \V{rb.cases} cases.
$B_{\mathrm{cert}}/(R - B_{\mathrm{cert}})$ is at most \V{rb.t1.max} at $\tau = 10^{-1}$ and \V{rb.t3.max} at $\tau =
10^{-3}$; at $\tau = 10^{-6}$, $B_{\mathrm{cert}} \ge R$ in all \V{rb.t6.over} cases. The certified bound is set by
$B_{\mathrm{cert}}$: the measured discrepancies $d$ are two orders below it, so ratios between discrepancies compare
computed solutions only.

\paragraph{Scope}
The certificate assumes IEEE binary64 and x87 long-double arithmetic with no overflow, underflow or invalid operation in
its own evaluation, which a script checks per case. It trusts no library exponential: its own long-double routine,
whose argument reduction is checked on every run of representable inputs (supplement), errs by less than
$167u_{ld}$. The route measurements and the certificate refer to the same reference vector, by its SHA-256 digest %
(\V{rb.shamatch} of \V{rb.shacases} cases). A verifier that shares no code with the certificate, trusting the
platform's exponential to one ulp, passes all \V{cv.cases} cases and gives $B_{\mathrm{cert}}/(R - B_{\mathrm{cert}})$
at most \V{cv.t1.max} at $\tau = 10^{-1}$. Enclosed: $\ell$, the pointwise bounds, $B_{\mathrm{cert}}$, $R$ and the
ratio~\eqref{eq:c3} at $\tau = 10^{-1}$ and $10^{-3}$. Not enclosed: the predictive variances, ratios between routes,
and the statistical adequacy of the covariance model.

\paragraph{The projection of the inverse error}
Why does an inverse error large in its own terms stay small in the predictions? Let $E = \widehat X - K^{-1}$ be the
error against the exact inverse. Each output sees it only through a projection: the mean through $k_j^TEb$, the
variance $v_j = c_0 - k_j^TK^{-1}k_j$ through $-k_j^TEk_j$. Relative to the variance, that error is bounded by
$\epsilon_X\,\alpha_j\,\kappa_j$, with $\epsilon_X = \|E\|_2/\|K^{-1}\|_2$, $\alpha_j =
\|K^{-1}\|_2\|k_j\|_2^2/(k_j^TK^{-1}k_j)$ the alignment the bound allows and $\kappa_j = k_j^TK^{-1}k_j/v_j$ the
cancellation in $v_j$. The measurements cannot use $E$ itself. They use $E_{\mathrm{ref}} = \widehat X -
X_{\mathrm{LAPACK}}$, the difference from LAPACK's computed inverse, with binary64 eigenvalue estimates for the norms,
and the variance discrepancies are taken against the variance from the refined solves, which the certificate does not
cover. They describe how computed results differ; they bound the exact-inverse quantities only up to those
references' own errors, which are not certified here. Measured on the H100 ($n = 4608$, location-major, 0.3 %
degrees), GEMMul8 with 14 moduli raises $\epsilon_X$ (with $E_{\mathrm{ref}}$) from
\V{pj.ckhlml3t1.4608.blkcublas.epsx} (native, blocked) to \V{pj.ckhlml3t1.4608.blkgemmul8m14.epsx} at $\tau =
10^{-1}$, but the realised fraction of the worst case, $|k_j^TE_{\mathrm{ref}}k_j|/(\|E_{\mathrm{ref}}\|_2\|k_j\|_2^2)$,
falls from \V{pj.ckhlml3t1.4608.blkcublas.varreal} to \V{pj.ckhlml3t1.4608.blkgemmul8m14.varreal} (medians). The
measured quadratic projections are small relative to the spectral-norm bound on the tested cross-covariance vectors;
since $E_{\mathrm{ref}}$ can be indefinite, the experiment does not distinguish weak alignment from cancellation within
those projections. The largest relative variance discrepancy
moves only from \V{pj.ckhlml3t1.4608.blkcublas.varrel} to \V{pj.ckhlml3t1.4608.blkgemmul8m14.varrel}, and at $\tau =
10^{-3}$ it is no larger than native's (\V{pj.ckhlml3t3.4608.blkgemmul8m14.varrel} against
\V{pj.ckhlml3t3.4608.blkcublas.varrel}). The cancellation factor $\kappa_j$ reaches \V{pj.ckhlml3t1.4608.kappav.max},
\V{pj.ckhlml3t3.4608.kappav.max} and \V{pj.ckhlml3t6.4608.kappav.max} at the three nuggets, and every route's variance
discrepancy grows with it; at $\tau = 10^{-6}$ every route, native ones included, shows large variance discrepancies
(\V{pj.ckhlml3t6.4608.cusolverpotri.varrel} for cuSOLVER).

\end{document}